\documentclass[reprint,twocolumn,superscriptaddress,prx,longbibliography,groupedaddress]{revtex4-2}

\usepackage{amsmath,amssymb}
\usepackage{graphicx}
\usepackage{dcolumn}
\usepackage{bm}
\usepackage{xr}
\usepackage[colorlinks=true]{hyperref}

\newcommand{\Wx}{\dot{W}_X}
\newcommand{\Wy}{\dot{W}_Y}
\newcommand{\Qx}{\dot{Q}_X}
\newcommand{\Qy}{\dot{Q}_Y}
\newcommand{\Wxy}{\dot{W}_{X\to Y}}
\newcommand{\Wyx}{\dot{W}_{Y\to X}}
\newcommand{\Ix}{\dot{I}_X}
\newcommand{\Iy}{\dot{I}_Y}
\newcommand{\Sigmax}{\dot{\Sigma}_X}
\newcommand{\Sigmay}{\dot{\Sigma}_Y}

\newcommand{\kB}{k_\mathrm{B}}

\newcommand{\dt}{\mathrm{d}t}
\newcommand{\mul}{\mu_\ell}
\newcommand{\mur}{\mu_\mathrm{r}}

\begin{document}

\title{Information Arbitrage in Bipartite Heat Engines}

\author{Matthew P.\ Leighton}
\thanks{These authors contributed equally.}
\author{Jannik Ehrich}
\thanks{These authors contributed equally.}
\author{David A.\ Sivak}%
\affiliation{Department of Physics, Simon Fraser University, Burnaby, BC, V5A 1S6, Canada.}%
\email{matthew\_leighton@sfu.ca}
\email{jehrich@sfu.ca}
\email{dsivak@sfu.ca}

\date{\today}

\begin{abstract}
Heat engines and information engines have each historically served as motivating examples for the development of thermodynamics. While these two types of systems are typically thought of as two separate kinds of machines, recent empirical studies of specific systems have hinted at possible connections between the two. Inspired by molecular machines in the cellular environment, which in many cases have separate components in contact with distinct sources of fluctuations, we study bipartite heat engines. We show that a bipartite heat engine can only produce net output work by acting as an information engine. Conversely, information engines can only extract more work than the work consumed to power them if they have access to different sources of fluctuations, i.e., act as heat engines. We illustrate these findings first through an analogy to economics and a cyclically controlled 2D ideal gas. We then explore two analytically tractable model systems in more detail: a Brownian-gyrator heat engine which we show can be reinterpreted as a feedback-cooling information engine, and a quantum-dot information engine which can be reinterpreted as a thermoelectric heat engine. Our results suggest design principles for both heat engines and information engines at the nanoscale, and ultimately imply constraints on how free-energy transduction is carried out in biological molecular machines.
\end{abstract}

\maketitle

\section{Introduction}
The last 200 years have seen thermodynamics evolve from its infancy in Carnot's ``Reflections on the motive power of fire"~\cite{carnot1824reflections} to being the dominant paradigm for studying how energy moves and changes in systems ranging from human-created machines~\cite{rajput2009engineering}, to large-scale astronomical structures like stars~\cite{leblanc2011introduction} and even the universe itself~\cite{weinberg2008cosmology}, as well as small-scale molecular machines operating within living cells~\cite{brown2019theory}. Historically, thermodynamics was developed to study the behavior and performance of heat engines. The quest to design a more efficient steam engine ultimately led to the formulation and refinement of the first and second laws of thermodynamics.

One of the current frontiers of thermodynamics lies in understanding how energy is transformed at microscopic scales. This is the domain of \emph{stochastic thermodynamics}~\cite{Jarzynski2011_Equalities,seifert2012stochastic,Peliti2021_book}, which also applies to the processes occurring inside living organisms far from equilibrium~\cite{brown2019theory}. In contrast to the classical thermodynamic arena of heat engines operating between different temperatures, biological processes are typically assumed to be isothermal, with fluctuations often treated as homogeneous and isotropic.

However, recent experimental and theoretical developments have shed light on possible departures from uniform fluctuations. For example, experiments suggest the mitochondrial temperature could be as much as $10$K hotter than the rest of the cell~\cite{Pinol2020_Real-Time,macherel2021conundrum,Wu2022_Intracellular,Di2022_Spatiotemporally}. This temperature difference could conceivably be accessed by the molecular machine ATP synthase which straddles the mitochondrial membrane. As another example, light-harvesting machines like photosystem~II~\cite{vinyard2013photosystem} are driven out of equilibrium by solar photons. Reactions induced by blackbody or monochromatic radiation can be thermodynamically treated as coupling to a heat bath at the blackbody temperature required to produce the observed intensity profile~\cite{buddhiraju2018thermodynamic,penocchio2021nonequilibrium,corra2022kinetic}. Lastly, the cellular interior supports a host of \emph{active fluctuations}~\cite{Mizuno2007_Nonequilibrium,Gallet2009_Power,pietzonka2019autonomous,dabelow2019irreversibility} powered by metabolic activity via the motion of large cytosolic components, for example enzymes and related complexes~\cite{parry2014bacterial,guo2014probing} or the cytoskeletal network~\cite{fodor2015activity}.

Such biological systems are typically composed of interacting degrees of freedom, which may thus be separately influenced by fluctuations of different strengths. The theory of bipartite stochastic thermodynamics~\cite{hartich2014stochastic,Diana2014_mutual,horowitz2014thermodynamics} describes energy and entropy balance at the level of individual subsystems, and allows for quantification of internal energy and information flows~\cite{Ehrich2023_Energy} between coupled subsystems. In such setups, different sources of fluctuations can be leveraged to improve performance. For example, active fluctuations speed \emph{in vitro} kinesin operation~\cite{ariga2021noise} and enzymatic catalysis~\cite{Tripathi2022_Acceleration}, and temperature differences increase output work in a model for ATP synthase~\cite{Grelier2023_Unlocking}. These effects are reminiscent of classical heat engines that alternately couple to different reservoirs at distinct temperatures. However, microscopic biological systems differ in that they are composed of interacting subsystems which each experience fluctuations from different sources. This motivates the study of two-component heat engines in which each part interacts with a heat bath at a different temperature. We call such systems \emph{bipartite heat engines}.

Studying small-scale systems also reveals the probabilistic nature of the laws of thermodynamics. This is illustrated by a famous thought experiment, known as Maxwell’s demon~\cite{Maxwell1867}: An intelligent being with microscopic information about the position and velocity of gas molecules can separate the fast from the slow ones, apparently violating the second law. “Exorcising” Maxwell’s demon~\cite{Leff2003_Maxwells} exposed a deep connection between information and thermodynamics~\cite{Parrondo2015_Thermodynamics}, specifically that information about a small fluctuating system can be exploited to perform useful tasks and that this information has a fundamental cost~\cite{landauer1961irreversibility,Parrondo2015_Thermodynamics}. Systems that leverage information to extract work are called \emph{information engines}, and there are abundant experimental realizations~\cite{Toyabe2010_Experimental,Camati2016_Experimental,Cottet2017_Observing,Masuyama2018_Information-to-work,Koski2014_Experimental_Realiz,Chida2017_Power,Admon2018_Experimental,Paneru2018_Losless,Admon2018_Experimental,Paneru2018_Optimal,Ribezzi2019_Large,Paneru2020_Efficiency,saha2021maximizing}.

Information engines fundamentally require a setup with two components: a \emph{controller} and a \emph{controlled system}. Information can then serve as a thermodynamic resource to make the controlled system, when considered on its own, do something seemingly forbidden by the second law, e.g., convert heat entirely into work. This comes at the controller’s expense because the apparent second-law violation entails an energetic cost~\cite{Ehrich2022_Energetic}---through Landauer’s principle~\cite{landauer1961irreversibility}---for performing feedback control and erasing previously acquired information.

Because of their interacting components and the relevance of (thermal) fluctuations, it is natural to ask whether molecular machines behave as information engines. Assuming a bipartite setup permits quantification of the information thermodynamics of such systems analogously to their energetics~\cite{crooks2019marginal}, by calculating an \emph{information flow}~\cite{allahverdyan2009thermodynamic,horowitz2020thermodynamic,Ehrich2023_Energy} quantifying the extent to which information is transduced through a composite system’s dynamics. This setup has been used to bound the dissipation of molecular sensors~\cite{Still2012_Prediction,Barato2014_Efficiency,Hartich2016_Sensory} and study the role information plays in bipartite molecular machines~\cite{large2021free,lathouwers2022internal,amano2022insights,Takaki2022_Information,leighton2023inferring,Grelier2023_Unlocking}. 

It has recently been suggested that information engines designed to leverage nonequilibrium fluctuations~\cite{paneru22,malgaretti22,Saha2022_Information} can greatly outperform their purely thermally driven counterparts. Output power can even surpass minimum control costs, rendering the information engine an energy harvester that operates between two reservoirs, the equilibrium thermal fluctuations affecting the controller and the nonequilibrium fluctuations affecting the controlled system. This setup is suggestively similar to that of a bipartite heat engine with a ``cold'' and a ``hot'' subsystem. In fact, like heat engines, such information engines are constrained by the Carnot bound, hinting at connections between the two engine types~\cite{feynman1965feynman,parrondo1996criticism,Still2020_cost_benefit_memory,Still2021_Partially}.

This paper elucidates a connection between heat engines exploiting the flow of heat from hot to cold, and information engines implementing Maxwell-demon-like exploitation of fluctuations. We broaden the perspective by illustrating that bipartite heat engines, where two engine components are each in contact with distinct reservoirs at different temperatures, can indeed be understood as information engines, and vice versa. To operate as a heat engine and extract energy from the temperature difference, the two components need to work together by exchanging entropy---in the form of information. This cooperation can be understood as an information engine in which the ``colder'' component acts as a controller that exploits the fluctuations of the ``hotter'' component. We illustrate our core findings using an analogy from economics, followed by a simple example consisting of a Carnot cycle on a 2D ideal gas with anisotropic temperature which we reveal to be a disguised information engine. 

Building on the theory of bipartite stochastic thermodynamics, we derive our most important mathematical result, Eq.~\eqref{eq:IFAR}: an inequality bounding the output work of a bipartite heat engine by the product of the temperature difference between the two reservoirs and an internal information flow between the two subsystems. This result, which we call the \textit{information-flow arbitrage relation} (IFAR), shows that a bipartite heat engine can only achieve net output work by supporting an information flow between its two subsystems, thus acting as an information engine. Conversely, a bipartite information engine can only produce more work than the energy cost required to run the controller when it operates between two temperatures, thus acting as a heat engine. This shift in perspective helps to establish the information engine as a useful mechanism for work extraction, and implies that any bipartite heat engine must implicitly contain this information-engine mechanism: Maxwell's demon lies at the heart of many real-world heat engines. 

Beyond elucidating connections between heat and information engines, our results also have applications for thermodynamic inference~\cite{seifert2019stochastic,leighton2023inferring}, for example providing a lower bound on the magnitude of information flow inside a bipartite heat engine. Calculations (detailed in Sec.~\ref{PSIIInference}) using numbers for photosystem II~\cite{vinyard2013photosystem} and bacteriorhodopsin~\cite{pinero2024optimization} suggest they support information flows as high as $10^3$\,bit/s. We verify these predictions by comparing with experimentally parameterized models.

The paper is organized as follows. In Sec.~\ref{sec:sec2} (developed using classical statistical physics without stochastic thermodynamics), we discuss classical and bipartite heat engines, illustrating our core findings first by analogy to economics and then with a worked example of a 2D ideal gas. Section~\ref{sec:bipartite_stoch_td} treats two-component heat engines using the stochastic thermodynamics of bipartite systems, and derives our central mathematical result, the information-flow arbitrage relation. Sec.~\ref{sec:models} quantitatively illustrates our results with two well-studied model systems for which all relevant quantities are analytically tractable: We uncover the information engine hidden in the Brownian gyrator (Sec.~\ref{sec:Brownian_gyrator}), and the heat engine in a quantum-dot Maxwell demon (Sec.~\ref{sec:quantum_dot}). We discuss the implications of our findings in Sec.~\ref{sec:discussion}.

\section{Heat engines are entropy arbitrageurs}
\label{sec:sec2}
Fundamentally, heat engines trade energy with heat reservoirs. They receive a certain amount of energy in the form of heat $Q_X$ from the hot reservoir, and give a lesser amount of heat $-Q_Y$ to the cold reservoir, withdrawing the energy difference as output work $-W$. For heat engines operating at steady state or under periodic driving, energy is conserved, so the work extracted equals the difference in heats. 

Flows of another quantity, \emph{entropy}, determine how input heat is split among output work and output heat. The input heat from the hot reservoir comes with an increase $\Delta_XS$ in the entropy of the heat engine, with the ratio of energy to entropy bounded by the temperature of the reservoir (multiplied by Boltzmann's constant $\kB$). Because of the lower temperature of the cold reservoir, getting rid of the entropy $-\Delta_YS$ only requires giving off a smaller amount of heat. 

\begin{figure}[tb]
    \centering
    \includegraphics[width = 1\linewidth]{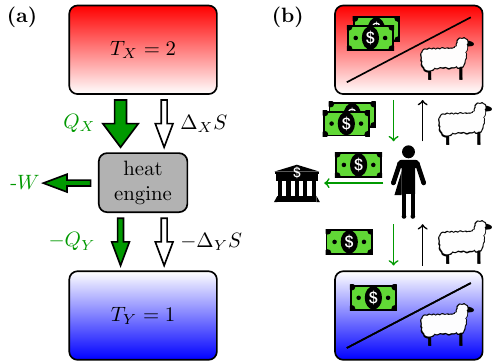}
    \caption{Heat engine as entropy arbitrageur. (a) Energy (green filled arrows) and entropy (unfilled arrows) flows in an ideal heat engine operating between a hot ($T_X=2$) and a cold ($T_Y=1$) reservoir. (b) Arbitrage analogy from economics. The heat engine is an arbitrageur buying sheep (entropy reduction) for a low price and selling sheep (entropy increase) for a higher price at a different market, pocketing the difference in money (energy).}
    \label{fig:arbitrageur}
\end{figure}

The greater the temperature difference, the less of the input energy needs to flow to the cold reservoir to maintain entropy balance, and thus the more energy can be extracted. Figure~\ref{fig:arbitrageur}(a) illustrates the energy and entropy flows in a heat engine in contact with two heat reservoirs at temperature $T_X=2$ and $T_Y=1$, respectively. For simplicity, we stick to a temperature ratio of $T_X/T_Y =2$ throughout this paper. We take the convention that work and heat flows into the system are positive.

The analysis invites an analogy from economics: When in two markets the prices for the same good differ, a market participant can make risk-free profit by buying from one market at a lower price and selling at another market for a higher price, pocketing the difference. This practice is \emph{arbitrage}~\cite{Samuelson2010_Economics} and people engaging in it are \emph{arbitrageurs}. The heat engine described above is such an arbitrageur: It ``buys'' a reduction in entropy from the cold reservoir for a smaller amount of energy and ``sells'' an equal entropy increase to the hot reservoir for a larger amount of energy, pocketing the difference. Inspired by~\cite{MinutePhysics}, we depict the analogy with an arbitrageur trading sheep at different markets in Fig.~\ref{fig:arbitrageur}(b). In this analogy money corresponds to energy, while sheep correspond to entropy reduction---since by the second law entropy cannot spontaneously decrease, decreasing entropy most cost something.

Trading sheep this way drives prices in the two markets, through the forces of supply and demand, toward equality, known as an \emph{arbitrage equilibrium}; aptly named, since in the thermodynamic context the heat flow through the engine eventually leads to an intermediate temperature in the reservoirs and a global \emph{thermodynamic equilibrium}. Just as processes in nature evolve towards an equilibrium, so do economic forces impel market participants to engage in arbitrage, pushing the market towards equilibrium. This analogy between heat engines and market arbitrage is not new~\cite{ellerman1984arbitrage,ellerman2000towards} and, though different from our approach here, has even been used in the context of an information engine~\cite{touzo2021information}. 

We consider the analogy particularly illuminating because in the same way in which a trader cannot increase the number of sheep by making them out of thin air (except by spending money) the second law forbids decreasing entropy (except by spending energy to do so). Hence, in both cases optimal efficiency is achieved by conserving the traded good. Inefficiencies in thermodynamic engines are expressed by unnecessary entropy production that increases the amount of heat that needs to be dissipated to break even. Similarly, if the arbitrageur somehow loses some sheep between the markets without compensation, they sell fewer sheep and make a smaller profit.

\subsection{Bipartite heat engines: Information arbitrage}
\label{sec:bipartiteheatengines}

\begin{figure*}[t!]
    \centering
    \includegraphics[width = 1\textwidth]{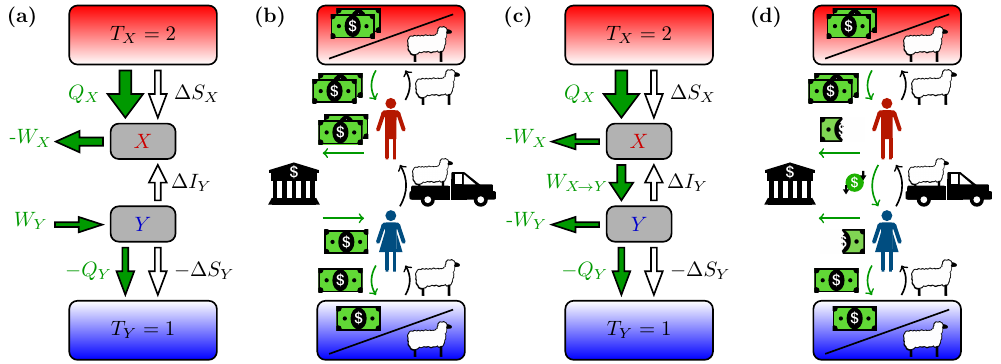}%
    \caption{Ideal bipartite heat engines and their economic analogs. (a) Energy (green filled arrows) and entropy (unfilled arrows) flows. The flow of entropy between engine components $X$ and $Y$ is conventionally identified as an information flow (in the opposite direction). In (a), $X$ and $Y$ cannot exchange energy, so $Y$ requires external input work. (b) Arbitrage analogy of (a): Two market participants must work together to realize the arbitrage scheme, requiring a flow of sheep from the lower-priced to the higher-priced market as well as a flow of money from the higher-priced market to the lower-priced market. (c) Same as (a) but with $X$ and $Y$ each extracting work. (d) Arbitrage analogy of (c).}
    \label{fig:arbitrageur_coop}
\end{figure*}

Not all engines are in simultaneous or alternating contact---as in conventional Carnot analysis---with two heat reservoirs. Instead, some engines are composed of distinct \emph{subsystems} that are each in permanent contact with a different reservoir at a given temperature. Macroscopic examples include thermoelectric devices, where two coupled junctions in contact with different temperatures collectively achieve work output~\cite{he2018thermoelectric}. Microscopic examples include the molecular machines mentioned in the Introduction, which are made up of different components that can be exposed to different sources of fluctuations from temperature gradients~\cite{Pinol2020_Real-Time,macherel2021conundrum,Wu2022_Intracellular,Di2022_Spatiotemporally}, hot thermal radiation~\cite{corra2022kinetic}, or active fluctuations~\cite{Mizuno2007_Nonequilibrium,Gallet2009_Power, parry2014bacterial,guo2014probing,fodor2015activity,ariga2021noise}.

Motivated by these examples, we now consider \emph{bipartite heat engines}, where two subsystems each interact with only one reservoir at a distinct temperature. Collectively, the subsystems can act as a heat engine, conducting heat from hot to cold and producing work output. Figure~\ref{fig:arbitrageur_coop}(a) depicts such a setup: The larger input heat $Q_X$ is entirely converted to output work $-W_X$, while the smaller input work $W_Y$ is entirely converted to output heat $-Q_Y$. Comparing with Fig.~\ref{fig:arbitrageur}(a), $-W = -W_X - W_Y >0$, i.e., net output work is positive. 

The setup portrayed in Fig.~\ref{fig:arbitrageur_coop}(a) does not require a flow of energy through the machine because heat and work are converted locally to satisfy energy balance in each subsystem; however, the machine requires a flow of entropy from the hot subsystem to the cold subsystem in order for each subsystem to satisfy a local second law. 

This entropy flow quantifies changes in the joint statistics of the degrees of freedom of the two subsystems. Specifically, entropy flows between subsystems when one subsystem, e.g.\ $X$, acts to increase the joint entropy by, for example, decreasing correlations between $X$ and $Y$, thereby decreasing their mutual information. Conversely, the dynamics of $Y$ can decrease the joint entropy by increasing correlations or more broadly mutual information. Thus in stochastic thermodynamics, the entropy transduction between components is called \emph{information flow}~\cite{allahverdyan2009thermodynamic,horowitz2014thermodynamics,Ehrich2023_Energy}. Section~\ref{sec:bipartite_stoch_td} gives a precise mathematical statement of information flow, but for now we make do with an intuitive explanation: In information theory, mutual information measures the mutual dependence between two variables. Information flow measures how much each subsystem tends to increase or decrease the mutual information between their statistical states. By the equivalence of information-theoretic (Shannon) entropy and thermodynamic entropy, subsystems may exchange thermodynamic entropy with each other by altering their mutual information.

This flow of information is the hallmark of an \emph{information engine} or Maxwell demon, which achieves conversion of heat to work using information, in apparent violation of the second law. In our setup we immediately see that the paradox results from only focusing on the $X$-subsystem in Fig.~\ref{fig:arbitrageur_coop}(a), which indeed transforms input heat to output work, and ignoring the $Y$-subsystem, which dissipates input work as output heat. 

We can also understand the necessity of a flow of entropy in terms of our previous arbitrage analogy, Fig.~\ref{fig:arbitrageur_coop}(b): Because each trader only has access to one market, they need to work together to make a net profit. Since they cannot produce new sheep (no decreasing entropy), a necessary requirement for their arbitrage scheme to work is that sheep are transported from the market with the lower exchange rate to that with the higher exchange rate.

Finally, bipartite heat engines may differ in which subsystem is capable of extracting work. For example, if the Y-subsystem has no access to a work source, the machine must transduce work from the hot to the cold side to ``pay for'' the entropy reduction of releasing heat to the cold reservoir. In terms of our analogy, in this case money must be transferred from one market participant to the other. Of course multiple different schemes of money transfer could be set up. For example, the two arbitrageurs could equally share profits, as illustrated in Figs.~\ref{fig:arbitrageur_coop}(c) and (d). Notice, however, that regardless of the details of the energy (money) extraction, entropy (sheep) must be transported from hot to cold reservoirs (lower- to higher-exchange-rate markets).

\subsection{Example: Carnot cycle for 2D ideal gas}
\label{sec:2dcarnotgas}

\begin{figure*}[tb]
\centering
\includegraphics[width = \textwidth]{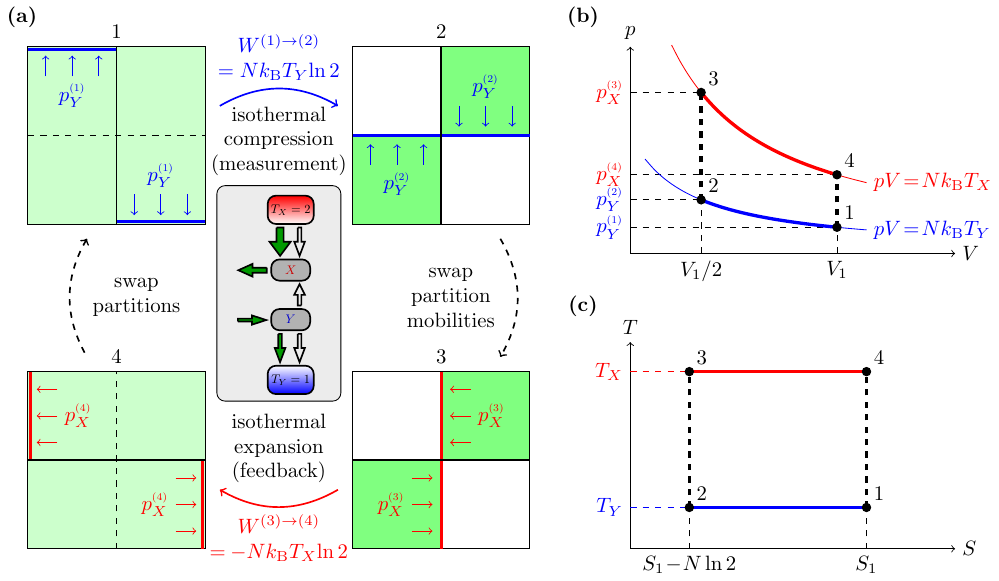}
\caption{Heat engine using a two-dimensional ideal gas in contact with heat baths acting in different directions. (a) Illustration of thermodynamic cycle. Small arrows indicate respective pressures in $X$ and $Y$ directions. Central inset: thermodynamic diagram showing presence/absence and directionalities of all energy, entropy, and information flows over the course of one cycle. (b) $p-V$ diagram indicating the pressures on the mobile partitions ($p$) and volumes ($V$) for each step in the cycle. (c) Corresponding $T-S$ diagram. In (b) and (c), red and blue curves denote isotherms at $T_X$ and $T_Y$, respectively, while black dashed lines denote isochores (or equivalently adiabats).}
\label{fig:ex_cycle}
\end{figure*}

If bipartite heat engines necessitate an information flow to generate net output power, one should be able to interpret a given bipartite heat engine in terms of an equivalent information engine. In this section we illustrate this principle in a slight modification of well-known thermodynamic process. Consider a 2D ideal gas, depicted in Fig.~\ref{fig:ex_cycle}(a), in which the position and momentum coordinates in the vertical direction are assumed to not interact with those in the horizontal direction, and the coordinates in the two directions are in contact with different thermal reservoirs at respective temperatures $T_X$ and $T_Y < T_X$. This idealized setup could be achieved, for example, with a monatomic gas cooled to low temperature in a container with perfectly elastic walls held at different temperatures. In the following we show how such a setup is harnessed to execute the thermodynamic cycle of a heat engine. Then, in Sec.~\ref{sec:info_engine_disguise} we interpret the thermodynamic process in terms of the well-known Szilard engine. Similar systems have been proposed and analyzed previously~\cite{gey2017exploiting,Still2020_cost_benefit_memory,Still2021_Partially}, but without considering connections between the two engine types.

The gas is confined to a square container of side length $L$ and exerts different pressures $p_X$ and $p_Y$ on the vertical and horizontal container walls, respectively. The ideal-gas law then gives relationships between the pressures $p_X$ and $p_Y$ (defined as forces per unit length), container volume $V=L^2$, and temperatures:
\begin{subequations}
\begin{align}
    p_X V &= N \kB T_X\,\\
    p_Y V &= N \kB T_Y\,,
\end{align}
\end{subequations}
where $N$ is the number of molecules.

Now consider the following thermodynamic process, depicted in Fig.~\ref{fig:ex_cycle}(a): 
\begin{enumerate}
    \item[Init.] The container is partitioned vertically into equal volumes. 
    \item[$1\to 2$] The gas is reversibly compressed in the $Y$-direction to half the volume, compressing from the top on the left side and from the bottom on the right. This compression creates a horizontal partition separating the top and bottom of the container.
    \item[$2\to 3$] The two parts of the horizontal partition are now fixed in place, while the vertical partition is split into two mobile parts which can move horizontally.
    \item[$3\to 4$] The gas is reversibly expanded to the original volume.
    \item[$4\to 1$] The horizontal partition is removed and 
    instantaneously
    replaced with the vertical partition.
\end{enumerate}

Figure~\ref{fig:ex_cycle}(b) shows the $p\mathrm{-}V$ diagram using at each step the relevant pressure $p_X$ or $p_Y$. The cycle contains two \emph{isothermal} steps and two \emph{isochoric} (constant volume) steps. The latter correspond to swapping the partition and thereby swapping which pressure is relevant for the gas expansion or contraction. As illustrated in Fig.~\ref{fig:ex_cycle}(c), this instantaneous swapping of the container partitions is not only isochoric but also \emph{adiabatic} since no heat is exchanged with the heat baths. This is possible because the engine essentially has two working media at different fixed temperatures: the $X$ and $Y$ components of the ideal gas. 

During the isochoric steps ($2\!\to\!3$ and $4\!\to\!1$), no heat is exchanged with the baths, and the compression ($1\!\to\!2$) and expansion ($3\!\to\!4$) steps are isothermal at respective temperatures $T_Y$ and $T_X$. Hence the work done on the gas during compression is $W_{1 \to 2} = N\kB T_Y \ln 2$ and during expansion is $W_{3 \to 4} = -N\kB T_X \ln 2$. Heat flows into the gas during the expansion, $Q_{3 \to 4} = N \kB T_X \ln 2$, and out of the gas during compression, $Q_{1 \to 2} = -N\kB T_Y \ln 2$. Hence, in one cycle the heat engine outputs net work
\begin{align}
    -W_{1 \to 2} - W_{3 \to 4} &= N\kB \left(T_X - T_Y \right) \ln 2 \label{eq:2D_gas_output_work}
\end{align}
at Carnot efficiency
\begin{align}
    \eta &= \frac{-W_{1 \to 2} - W_{3 \to 4}}{Q_{3 \to 4}} = 1 - \frac{T_Y}{T_X}\,.
\end{align}

\subsection{Szilard engine in disguise}\label{sec:info_engine_disguise}
There is an information engine hidden in this heat engine. Imagine only having access to the $X$-position of each gas molecule during the cycle: In step $2\!\to\!3$ the container is divided and each gas molecule is either left or right of the partition. In step $3\!\to\!4$ input heat drives an isothermal expansion, and work is extracted from the gas as the piston is moved across the correct respective empty half of the container (appearing to act on hidden knowledge of the system state): It seems as if heat is entirely converted into work. However, with full access to the state space we recognize that this expansion step, seemingly requiring hidden knowledge, is preceded by a compression step that correlates each molecule's $Y$~position with its $X$~position, and the expansion step utilizes the $Y$ position as a memory to execute the expansion into the correct half. 

Imagining this process with a single molecule, we obtain the famous Szilard engine~\cite{Szilard1929,lutz2015information}, arguably the simplest information engine used to illustrate the Maxwell-demon paradox~\cite{Leff2003_Maxwells}. For the Szilard engine, the paradox is resolved by explicitly accounting for the memory degree of freedom and using Landauer's principle~\cite{landauer1961irreversibility} to show that information erasure carries a thermodynamic cost. In analogy to this, we analyze the information thermodynamics~\cite{Parrondo2015_Thermodynamics} of the interplay between the $X$ and $Y$ components of the gas.

In the initial, uncompressed state the $N$-molecule gas has joint entropy
\begin{align}
    S_1[X,Y] = N\ln V + \frac{N}{2}\ln{T_X} + \frac{N}{2}\ln{T_Y} \ ,
\end{align}
up to constants irrelevant to the analysis. (Note that while we appear to take logarithms of dimensional quantities; this is because we have omitted additional constant factors~\cite{callen1998thermodynamics}.) Using the container's side length $L$ (and hence volume $V=L^2$), the joint entropy is
\begin{subequations}
\begin{align}
    S_1[X,Y] &= S_1[X] + S_1[Y]\,,
\end{align}
for marginal entropies
\begin{align}
    S_1[X] &= N\ln L + \frac{N}{2} \ln T_X\\
    S_1[Y] &= N\ln L + \frac{N}{2} \ln T_Y\,.
\end{align}
\end{subequations}
Initially, at equilibrium without any coupling, the subsystems $X$ and $Y$ are independent and hence they have zero mutual information: $I_1[X;Y] = S_1[X] + S_1[Y] - S_1[X,Y]=0$.

The compression step $1\!\to\!2$ does not change the one-dimensional phase space available to each individual component. Consequently, the marginal entropies remain unchanged: $S_2[X] = S_1[X]$ and $S_2[Y] = S_1[Y]$. Nonetheless, the joint entropy has been reduced by $N\ln 2$, since the (joint) phase-space volume available to each molecule has been halved. Since the joint entropy has been reduced while holding the marginal entropies constant, mutual information has been introduced between the components:
\begin{subequations}
\begin{align}
    S_2[X,Y] &= S_2[X] + S_2[Y] - I_2[X;Y]\\
    &= S_1[X,Y] - N\ln 2\,,
\end{align}
\end{subequations}
and hence
\begin{align}
    I_2[X;Y] = N \ln 2\,. \label{eq:mutual_info_gas}
\end{align}
This can be thought of as using a binary memory variable $Y$ to encode each molecule's coarse-grained $X$-position: whether it is left or right of the divider. Therefore each molecule's coarse-grained $Y$-position (\emph{above} or \emph{below} the divider) acts as a memory of each molecule's coarse-grained $X$-position. Reducing entropy incurs a thermodynamic cost, expressed by the work done (or equivalently the heat flow) in this step,
\begin{align}
    W_{1 \to 2} = -Q_{1 \to 2} = \kB T_Y I_2[X;Y] \,. \label{eq:2d_gas_W12}
\end{align}
Hence the work $W_{1 \to 2}$ produces mutual information $I_2[X;Y]$.

Swapping the mobilities of the partitions leaves the entropy unaffected: $S_3[X;Y] = S_2[X,Y]$. The expansion step $3\!\to\!4$ returns the engine back to the initial state and thus
\begin{align}
    S_4[X,Y] = S_1[X,Y]
    = S_1[X] + S_1[Y]\,.
\end{align}
Therefore, the expansion ``uses up'' the previously created mutual information $I_2[X;Y]$ to do work:
\begin{align}
    W_{3 \to 4} = - \kB T_X I_2[X;Y]\,. \label{eq:2d_gas_W34}
\end{align}

From this point of view, the engine cycle consists of a measurement step $1\!\to\!2$ creating information, and feedback step $3\!\to\!4$ exploiting that information. Thereby the $Y$-component of the gas acts as the controller and memory of an information engine whose working substance is the $X$-component of the gas.

\section{Bipartite stochastic thermodynamics and information-flow arbitrage relation}\label{sec:bipartite_stoch_td}
We now mathematically formalize the above ideas to demonstrate that any bipartite heat engine must utilize information to achieve positive output power. Specifically, we derive an inequality that bounds the output power of a bipartite heat engine by its information flow.

We turn to the theory of bipartite stochastic thermodynamics~\cite{hartich2014stochastic,Diana2014_mutual,horowitz2014thermodynamics}, where a single stochastic system is decomposed into two distinct subsystems (respectively characterized by degrees of freedom $X$ and $Y$). These two degrees of freedom evolve according to coupled stochastic equations of motion (either discrete or continuous), exchanging energy and entropy both among themselves and with their environments. Throughout this article, we refer to the two parts of a bipartite system as either \emph{subsystems} (when discussing their thermodynamics), \emph{degrees of freedom} (when discussing their dynamics), or \emph{components} (when discussing specific realizations of such systems).

This formalism has been developed primarily to study two-component molecular machines such as ATP synthase~\cite{lathouwers2020nonequilibrium,leighton2023inferring}, transport motors pulling cargo~\cite{zimmermann2015effective,brown2019pulling,leighton2022performance,leighton2023inferring}, synthetic molecular motors~\cite{amano2022insights}, and light-driven pumps~\cite{corra2022kinetic}. Other model systems include information engines~\cite{horowitz2014second,horowitz2014thermodynamics,barato2017thermodynamic,Ehrich2022_Energetic}, cellular sensors~\cite{Still2012_Prediction,Barato2014_Efficiency,Hartich2016_Sensory}, and coupled quantum dots~\cite{horowitz2014thermodynamics}.

A key feature of bipartite stochastic thermodynamics is the ability to quantitatively identify flows of energy and entropy for two coupled subsystems. In addition to the global second law, we can formulate subsystem-specific second laws describing entropy balance at the level of individual subsystems. This decomposition is most simply performed by assuming that the dynamics of the two degrees of freedom are \emph{bipartite}~\cite{hartich2014stochastic,Diana2014_mutual,horowitz2014thermodynamics}. For continuous dynamics this means that the noise sources for the two degrees of freedom are statistically independent, while for discrete dynamics this means the two degrees of freedom cannot simultaneously change states. The bipartite assumption does not preclude energetic coupling between the two subsystems, allowing them to explicitly exchange energy~\cite{fogedby2017minimal,amano2022insights}. While stochastic thermodynamics can be formulated for two-component systems without the bipartite assumption~\cite{chetrite2019information}, it remains near-universal in the field. The bipartite assumption is especially appropriate for systems in contact with distinct baths at different temperatures, where by construction the noise sources acting on the two subsystems originate from interactions with different thermal reservoirs.

Subsystem-specific second laws quantify the respective rates of entropy production $\Sigmax$ and $\Sigmay$ due to the  dynamics of $X$ and $Y$:
\begin{subequations}\label{eq:secondlaws}
\begin{align}
\Sigmax & = \mathrm{d}_tS[X] - \beta_X\Qx -\Ix\geq 0,\label{eq:Xsecondlaw}\\
\Sigmay & = \mathrm{d}_tS[Y] - \beta_Y\Qy - \Iy\geq 0.\label{eq:Ysecondlaw}
\end{align}
\end{subequations}
Here $\mathrm{d}_tS[X]$ ($\mathrm{d}_tS[Y]$) is the rate of change of the marginal entropy of $X$ ($Y$), $\Qx$ ($\Qy$) is the rate of heat into $X$ ($Y$), $\Ix$ ($\Iy$) is the information flow due to $X$ ($Y$), and $\beta_X\equiv (\kB T_X)^{-1}$ ($\beta_Y\equiv(\kB T_Y)^{-1}$). A recent review article~\cite{Ehrich2023_Energy} provides further details of how Eqs.~\eqref{eq:secondlaws} are derived for stochastic systems.

The information flows $\Ix$ and $\Iy$ describe the respective rates of change of the mutual information $I(X;Y)$ due to dynamics of $X$ and $Y$. These flows have many equivalent definitions; we find the most illuminating to be:
\begin{subequations}\label{eq:infoflowdefs}
\begin{align}
\Ix & \equiv \lim_{\dt\to 0} \frac{I[X(t+\dt);Y(t)] - I[X(t);Y(t)]}{\dt}, \label{eq:def_X^info_flow}\\
\Iy & \equiv \lim_{\dt\to 0} \frac{I[X(t);Y(t+\dt)] - I[X(t);Y(t)]}{\dt}. \label{eq:def_Y^info_flow}
\end{align}
\end{subequations}
These definitions, which hold for both continuous and discrete degrees of freedom, equate the information flows with the instantaneous change in the mutual information when one coordinate varies with the other held fixed. The sum of the two information flows gives the time derivative of the mutual information,
\begin{equation}
\mathrm{d}_t I[X;Y] = \Ix + \Iy.
\end{equation}

In addition to the second laws~\eqref{eq:secondlaws}, bipartite systems also satisfy a first law describing energy balance:
\begin{equation}\label{eq:firstlaw}
\dot{W} + \Qx + \Qy = \mathrm{d}_tE.
\end{equation}
Here $E$ is the internal energy of the system, and $\dot{W}$ is the rate of work into the system, which in general may include contributions from both nonconservative driving forces and changes in potential energy due to varying external control parameters.

\subsection{Nonequilibrium steady states}\label{sec:ness}
We now consider the special case of autonomous systems not subject to time-dependent external control. Energy balance at the level of individual subsystems is quantified by local first laws:
\begin{subequations}\label{eq:firstlaws}
\begin{align}
\Wx + \Qx & = \Wxy\,,\\ 
\Wy + \Qy & = \Wyx\,,
\end{align}
\end{subequations}
where $\Wx$ ($\Wy$) is the rate of work into $X$ ($Y$) due to nonequilibrium driving forces, and $\Wxy$ ($\Wyx$) is the transduced power from $X$ to $Y$ ($Y$ to $X$). The latter definition arises from treating $X$ ($Y$) as a control parameter performing work on $Y$ ($X$)~\cite{Ehrich2023_Energy}.

At nonequilibrium steady state (NESS), the first and second laws simplify. The marginal entropies, the mean system energy, and mutual information all remain constant:
\begin{subequations}\label{eq:steadystateconditions}
\begin{align}
\mathrm{d}_tS[X] =0 & =\mathrm{d}_tS[Y],\\
0& =\mathrm{d}_tE=\Wxy+\Wyx,\\
0& =\mathrm{d}_tI=\Ix+\Iy.
\end{align}
\end{subequations}
The first and second laws combine to yield two inequalities constraining sums of the external work rates, the transduced power, and the information flow:
\begin{subequations}\label{eq:firstsecondlaws}
\begin{align}
\Sigmax & = \beta_X\Wx +\beta_X\Wyx+\Iy\geq0\,,\\
\Sigmay & = \beta_Y\Wy -\beta_Y\Wyx -\Iy\geq0\,.
\end{align}
\end{subequations}

Multiplying Eqs.~\eqref{eq:firstsecondlaws} by $\kB$ and the respective subsystem temperatures $T_X$ and $T_Y$, summing the two equations, and rearranging yields an upper bound on the total output power in terms of the temperature difference and the information flow:
\begin{equation}\label{eq:IFAR}
-\dot{W}= -\Wx-\Wy \leq \kB(T_X-T_Y)\Iy.
\end{equation}
This first main result, which we call the \emph{information-flow arbitrage relation} (\emph{IFAR}), holds significant implications for the functioning of bipartite heat engines. It makes precise the notion, described in Sec.~\ref{sec:bipartiteheatengines} through analogy to economic arbitrage, that information flows are necessary; to achieve net output work ($-\dot{W}>0$), there must be a flow of information between the two subsystems. Moreover, it must be the colder subsystem whose dynamics increase correlations (and thus the hotter subsystem whose dynamics reduce them): if $T_X>T_Y$, then a functional heat engine must have $\Iy>0$ (and thus $\Ix<0$).

In addition to showing that bipartite heat engines require information flows, the IFAR also provides a necessary condition for information engines to achieve net output power $-\dot{W}>0$: a positive information flow $\Iy$ is not sufficient; a temperature difference is also required. Only then does the controller at $T_Y$ ``pay" less energy to create correlations than the controlled system at $T_X>T_Y$ extracts by consuming those correlations. This reflects a fundamental connection between information engines and bipartite heat engines: to achieve net output power, an information engine must leverage fluctuations of different strengths (e.g., a temperature difference, which is the driving force for a heat engine), while a bipartite heat engine must contain an information flow (the hallmark of an information engine).

While the IFAR bounds the net output power $-\dot{W}$, with dimensions of energy divided by time, this output power is spread across two different temperatures. These different temperatures constitute different conversion rates between energy and entropy, and so the two output powers $-\Wx$ and $-\Wy$ are each measured relative to different thermal backgrounds. We can take these different ``exchange rates" into account by considering the sum of the two output powers scaled by their respective temperatures, $-\beta_X\Wx-\beta_Y\Wy$. Similar to the derivation of the IFAR, summing Eqs.~\eqref{eq:firstsecondlaws} and rearranging yields
\begin{equation}\label{eq:TPAR} 
-\beta_X\Wx-\beta_Y\Wy \leq (\beta_X-\beta_Y)\Wyx,
\end{equation}
which we call the \emph{transduced-power arbitrage relation} (\emph{TPAR}). Our second main result, the TPAR states that to obtain a positive sum of scaled output powers, the hot ($X$) subsystem must transduce work to the cold ($Y$) subsystem so that $\Wyx<0$.

An intriguing possible configuration of a bipartite heat engine is one where net output work is extracted from both subsystems (both $\Wx<0$ and $\Wy<0$), as illustrated in Fig.~\ref{fig:arbitrageur_coop}(c). (We will show an explicit example of such a heat engine in section~\ref{sec:Brownian_gyrator}.) Combining the IFAR~\eqref{eq:IFAR} and TPAR~\eqref{eq:TPAR} shows that extracting net output work from both subsystems is only possible for specific directions of the internal energy and information flows. In particular, transduced work must flow from hot to cold ($\Wyx<0$) while information must flow from cold to hot ($\Iy>0$).

\subsection{Interpretation in terms of environmental potentials}\label{environment_potential_interpretations}
For an alternative interpretation of these results, consider the thermodynamics from the perspective of the environment, comprised of two equilibrium thermal reservoirs at $T_X$ and $T_Y$ respectively, along with two work reservoirs. The portions of the environment in contact with $X$ and $Y$ are assumed to each have well-defined energies ($U_X^\mathrm{env}$ and $U_Y^\mathrm{env}$) and entropies ($S_X^\mathrm{env}$ and $S_Y^\mathrm{env}$). From these state functions we can construct thermodynamic potentials for the environment, for example the Helmholtz free energy
\begin{equation}\label{eq:env_feng}
F_\mathrm{env} \equiv U_X^\mathrm{env} - \kB T_X S_X^\mathrm{env} + U_Y^\mathrm{env} - \kB T_YS_Y^\mathrm{env},
\end{equation}
and free entropy (also known as the Massieu potential~\cite{callen1998thermodynamics})
\begin{equation}\label{eq:env_fent}
\Phi_\mathrm{env} \equiv S_X^\mathrm{env} - \beta_XU_X^\mathrm{env} + S_Y^\mathrm{env} - \beta_YU_Y^\mathrm{env}.
\end{equation}
Since the reservoirs interact (and thus exchange energy and entropy) only with their respectively coupled subsystems, the rates of change of their energies and entropies are $\dot{U}_X^\mathrm{env} = -\Wx -\Qx$ and $\dot{S}_X^\mathrm{env} = -\beta_X\Qx$, and likewise for $\dot{U}_Y^\mathrm{env}$ and $\dot{S}_Y^\mathrm{env}$. We can then compute the rates of change of the two environmental potentials when the system is at steady state:
\begin{subequations}
\begin{align}
\dot{F}_\mathrm{env} & = -\Wx - \Wy,\\
\dot{\Phi}_\mathrm{env} & = \beta_X\Wx + \beta_Y\Wy.
\end{align}
\end{subequations}
We then substitute these definitions into the left hand sides of the IFAR~\eqref{eq:IFAR} and TPAR~\eqref{eq:TPAR} to reformulate them in terms of rates of change of environmental potentials:
\begin{subequations}
\begin{align}
\dot{F}_\mathrm{env} & \leq \kB (T_X-T_Y)\Iy,\\
-\dot{\Phi}_\mathrm{env} & \leq (\beta_X-\beta_Y)\Wyx.
\end{align}
\end{subequations}
These reformulations lead to a new interpretation with a pleasing symmetry. The IFAR states that the rate at which the system can leverage a temperature difference to increase the free energy of the environment is limited by the rate of internal information (entropy) transduction, while conversely the TPAR states that the rate at which the system can decrease the free entropy of the environment is limited by the rate of internal energy transduction.

Intuitively, these results follow from the definitions of environmental free energy~\eqref{eq:env_feng} and free entropy~\eqref{eq:env_fent}. Since the energetic terms of the free energy~\eqref{eq:env_feng} are not modulated by temperature, and at steady state their sum must remain constant, it follows that $F_\mathrm{env}$ can only be increased by moving entropy from the hotter reservoir to the colder one. The reservoirs only interact indirectly via the system, which must then serve as a conduit for the entropy flow, which we call an information flow. Thus environmental free energy can only increase if the system supports an internal information flow. 

Conversely, the entropic terms of the free entropy~\eqref{eq:env_fent} are not modulated by temperature, and at steady state cannot decrease (by the second law), so it follows that $\Phi_\mathrm{env}$ can only be decreased by moving energy from the hotter reservoir to the colder one. As with entropy, energy can only be exchanged using the system as a conduit. Thus decreasing the environmental free entropy requires an internal energy flow, i.e., a transduced power from hot to cold. 

This interpretation of the two arbitrage relations is intimately connected with the economics analogy outlined earlier in Sec.~\ref{sec:sec2}. A bipartite heat engine which increases the free energy of the environment corresponds to a pair of arbitrageurs who cooperate to extract net money from two markets with different exchange rates; as illustrated in Fig.~\ref{fig:arbitrageur_coop} this is only possible when the two arbitrageurs exchange sheep (i.e., when the heat engine supports an information flow). Conversely, decreasing the free entropy of the environment corresponds to the arbitrageurs obtaining a net return of sheep, which in turn requires them to exchange money with each other (corresponding to the two subsystems supporting a transduced power).

\subsection{Connection to Carnot bound}
The IFAR~\eqref{eq:IFAR} relates the output work of a bipartite heat engine to the information flow and temperature difference. The output working of a heat engine operating between two reservoirs was famously first upper-bounded by Carnot~\cite{carnot1824reflections}, in terms of the temperature ratio and the input heat. Because in a bipartite heat engine, each of the two reservoirs is coupled to a distinct individual subsystem, the input heat from the ``hot" reservoir can only flow into the ``hot'' subsystem $X$. Then, the Carnot bound for heat engines at steady state is derived from the first law~\eqref{eq:firstlaw} (with $\mathrm{d}_t E=0$), and the global second law,
\begin{equation}\label{eq:globalsecondlaw}
\dot{\Sigma} = -\beta_X\Qx - \beta_Y\Qy\geq 0.
\end{equation}
This is simply a sum of the two subsystem-specific second laws~\eqref{eq:secondlaws}. Rearranging Eq.~\eqref{eq:globalsecondlaw} to get an upper bound on $\Qy$ and inserting into Eq.~\eqref{eq:firstlaw} yields the Carnot bound on the ouput power of a heat engine:
\begin{equation}\label{eq:carnot}
-\dot{W}\leq \left(1-\frac{T_Y}{T_X}\right)\Qx.
\end{equation}
The Carnot bound essentially states that in a heat engine input heat limits output work, with a proportionality constant dependent on the temperature ratio.

Now notice that we have $\dot Q_X \geq 0$, i.e., heat flows into subsystem $X$ (which, recall, is coupled to only one reservoir), something seemingly forbidden by the second law and the hallmark of a Maxwell demon~\cite{strasberg2013thermodynamics,Koski2015_On-chip,Ciliberto2020_Autonomous_demon,Freitas2021_Characterizing}. Under steady-state conditions~\eqref{eq:steadystateconditions}, the second law for the $X$ subsystem~\eqref{eq:Xsecondlaw} can be rewritten to bound the achievable input heat by the information flow,
\begin{equation}\label{eq:X^demon}
\Qx \leq \kB T_X \Iy.
\end{equation}
Inserting this inequality into the Carnot bound yields the information-flow arbitrage relation:
\begin{equation}
\underbrace{\lefteqn{\overbrace{\phantom{-\dot{W} \leq \left(1-\frac{T_Y}{T_X}\right)\Qx \;}}^{\text{Carnot}}}  -\dot{W} \leq \underbrace{\left(1-\frac{T_Y}{T_X}\right)\Qx\leq \kB (T_X-T_Y)\Iy }_{\text{2nd Law for }X}}_{\text{IFAR}}.
\end{equation}

Thus we find that the IFAR is in general looser than the Carnot bound on heat-engine output work. Nonetheless, IFAR broadens the perspective by showing that working bipartite heat engines necessarily require an information flow. Note also that IFAR is saturated for heat engines at equilibrium (where output work vanishes), as is the Carnot bound~\cite{brandner2015thermodynamics}.

\subsection{Periodic driving} \label{sec:periodicSS}
While biological systems of interest operate autonomously and are typically in nonequilibrium steady states, many human-engineered systems (e.g., classical heat engines and the 2D ideal gas considered in Sec.~\ref{sec:2dcarnotgas}) are controlled by periodic driving protocols. Likewise, experimental~\cite{Toyabe2010_Experimental,Camati2016_Experimental,Cottet2017_Observing,Masuyama2018_Information-to-work,Koski2014_Experimental_Realiz,Chida2017_Power,Admon2018_Experimental,Paneru2018_Losless,Admon2018_Experimental,Paneru2018_Optimal,Ribezzi2019_Large,Paneru2020_Efficiency,saha2021maximizing} and theoretical models~\cite{Bauer2012_Efficiency,Um2015_Total,Schmitt2015_Molecular,Bechhoefer2015,Lucero2021_Maximal,Still2021_Partially} of information engines typically use repeated feedback loops~\cite{Cao2009_Thermodynamics,Ponmurugan2010_Generalized,Horowitz2010_Nonequilibrium,Sagawa2012_Nonequilibrium,Sagawa2013_Role,Ehrich2017_Stochastic,crooks2019marginal} which comprise measurement, feedback, and relaxation steps. These setups can also be understood in terms of a periodic driving protocol that achieves the desired feedback~\cite{Ehrich2022_Energetic}. Information engines with access to a temperature difference~\cite{Still2020_cost_benefit_memory,Still2021_Partially} or nonequilibrium fluctuations that only affect their working medium and not their controller~\cite{paneru22,malgaretti22,Saha2022_Information} can have positive net work output. Here we show that the IFAR~\eqref{eq:IFAR} also holds for this class of systems.

For bipartite systems in periodic steady states, the IFAR is derived analogously to systems at NESS. We take as a starting point the instantaneous second laws~\eqref{eq:secondlaws}, and integrate over the cycle time $\tau$ to yield their cyclic counterparts:
\begin{subequations}\label{eq:periodicsecondlaws}
\begin{align}
\Sigma_X & = \Delta S[X] -\beta_XQ_X - \Delta I_X\geq0,\\
\Sigma_Y & = \Delta S[Y] -\beta_YQ_Y - \Delta I_Y\geq0.
\end{align}
\end{subequations}
Here $Q_X \equiv \int_0^\tau \Qx(t) \, \dt$, $\Sigma_X \equiv \int_0^\tau \Sigmax(t) \dt$, $\Delta S[X] \equiv S_\tau[X]-S_0[X]$, and $\Delta I_X \equiv \int_0^\tau \Ix \dt$, with analogous definitions holding for flows due to $Y$. Likewise, the global first law also integrates to yield
\begin{equation}\label{eq:periodicfirstlaw}
W + Q_X + Q_Y = \Delta E,
\end{equation}
where $W\equiv\int_0^\tau \dot{W}\dt$ and $\Delta E\equiv E(\tau)-E(0)$. 

For systems at periodic steady states, we require that all state variables are identical at the beginning and end of the cycle. Specifically these are the mean internal energy [$\Delta E = E(\tau) - E(0) = 0$], the two marginal entropies ($\Delta S[X] = S_\tau[X] - S_0[X]=0$ and $\Delta S[Y] = S_\tau[Y] - S_0[Y]=0$), and the mutual information ($\Delta I_X + \Delta I_Y = I_\tau[X;Y] - I_0[X;Y] =0$). Using these invariants to simplify the first~\eqref{eq:periodicfirstlaw} and second laws~\eqref{eq:periodicsecondlaws}, we derive the IFAR for periodically driven systems:
\begin{equation}
    -W \leq \kB(T_X-T_Y)\Delta I_Y. \label{eq:IFAR_periodic}
\end{equation}
The TPAR~\eqref{eq:TPAR} does not extend as easily to periodic steady states, since in the presence of external control, defining external work at the subsystem level requires a more nuanced analysis beyond the scope of this paper.

\subsection{Revisiting the 2D ideal-gas engine}
Having derived the IFAR for periodically driven systems, we return to the 2D ideal-gas engine in Sec.~\ref{sec:2dcarnotgas} to illustrate this relation.

Intuitively, the total $Y$-information flow over one cycle should be given in terms of the information acquired during the compression ($1\!\to\!2$) step, when the molecules' $Y$-coordinates change. To see why this is indeed true, consider first the \emph{coarse-grained} variables $\bar X^i~:=~\mathrm{sgn}(X^i)$ and $\bar Y^i := \mathrm{sgn}(Y^i)$ that respectively indicate whether a given molecule $i$ is left or right of and above or below the box's center. The gas molecules' coordinates are in equilibrium hence the conditional probability of the specific position given the coarse-grained position is uniform and factorizes,
\begin{equation}
p(X^i,Y^i|\bar X^i, \bar Y^i)=p(X^i|\bar X^i)\,  p(Y^i|\bar Y^i).
\end{equation}
This implies for the conditional entropy of the specific position given the coarse-grained position, $S[X^i,Y^i|\bar X^i, \bar Y^i]~=~S[X^i|\bar X^i] + S[Y^i|\bar Y^i]$\,. Therefore, the mutual information between the coordinates $X^i$ and $Y^i$ equals the mutual information between the coarse-grained coordinates $\bar X^i$ and $\bar Y^i$,
\begin{subequations}\label{eq:mutualinfoequiv}
\begin{align}
    I[X^i;Y^i] &= S[X^i] + S[Y^i] - S[X^i,Y^i]\\
    &=S[X_i,\bar{X}_i] + S[Y^i, \bar Y^i] - S[X^i,Y^i,\bar X^i, \bar Y^i]\label{eq:mutual_info_equal_2}\\ 
    &= S[\bar X^i] + S[\bar Y^i] - S[\bar X^i, \bar Y^i] \label{eq:mutual_info_equal_3}\\
    &\quad + \underbrace{S[X^i|\bar X^i] + S[Y^i|\bar Y^i] - S[X^i,Y^i|\bar X^i, \bar Y^i]}_{=0}\nonumber \\
    &= I[\bar X^i; \bar Y^i]\,,
\end{align}
\end{subequations}
where to get~\eqref{eq:mutual_info_equal_2} we used the fact that once a molecule's true position $(X^i,Y^i)$ is known, the coarse-grained position $(\bar X^i, \bar Y^i)$ is redundant information. From Eqs.~\eqref{eq:mutualinfoequiv}, it follows that to calculate information flow, we can replace all mutual information terms by their coarse-grained counterparts.

The total $Y$-information flow is
\begin{subequations}
\begin{align}
    \Delta I_Y &= \int_1^2 \dt\, \dot I_Y + \int_3^4 \dt\, \dot I_Y \label{33a}\\
    &= \int_1^2 \dt \lim_{\dt \to 0} \frac{I[\bar X(t+\dt);\bar Y(t+\dt)] - I[\bar X(t);\bar Y(t)]}{\dt} \nonumber \\
    &\;+ \int_3^4 \dt\, \lim_{\dt \to 0} \frac{I[\bar X(t);\bar Y(t)] - I[\bar X(t);\bar Y(t)]}{\dt}\label{33b}\\
    &= I_2[\bar X;\bar Y] - \underbrace{I_1[\bar X;\bar Y]}_{=0} = I_2[X;Y]\,,
\end{align}
\end{subequations}
where $I_2[X;Y]$ is the mutual information between the molecules' $X$ and $Y$ coordinates after compression, see Eq.~\eqref{eq:mutual_info_gas}. To get Eq.~\eqref{33b} we used the definition of the Y-information flow~\eqref{eq:def_Y^info_flow} and the fact that $\bar X^i(t+ \mathrm \dt)=\bar X^i(t)$ in step $1\!\to\!2$ and $\bar Y^i(t+ \mathrm \dt)=\bar Y^i(t)$ in step $3\!\to\!4$, i.e., during compression (expansion) the coarse-grained $X$($Y$)-coordinate does not change. Similarly, 
\begin{align}
    \Delta I_X = I_4[\bar X;\bar Y] - I_3[\bar X;\bar Y] =-I_2[X;Y]=-\Delta I_Y\,.
\end{align}
With the total net output work in Eq.~\eqref{eq:2D_gas_output_work} and the mutual information $I_2[X;Y] = N \ln 2$~\eqref{eq:mutual_info_gas}, we verify that the IFAR~\eqref{eq:IFAR_periodic} holds as an equality, which can be attributed to the fact that all steps in the cycle are carried out reversibly.

\subsection{Entropy arbitrage without bipartite structure}
\label{section:gear_derivation}
Thus far we have focused on bipartite heat engines composed of two distinct subsystems each in contact with a different thermal reservoir. In some cases however, it may not be possible to resolve distinct degrees of freedom corresponding to different subsystems, while still being able to ascribe the rates of different transitions to coupling with distinct reservoirs. An example would be a system described by a discrete set of states $\{z_i\}$, whose dynamics follow the master equation
\begin{equation}
\frac{\partial}{\partial t} p_i = \sum_j \left[\left(R_{ij}^A + R_{ij}^B\right)p_j - \left(R_{ji}^A + R_{ji}^B\right)p_i\right] \ ,
\end{equation}
for transition rates $R_{ij}^A$ and $R_{ij}^B$ respectively coupled to distinct thermal reservoirs $A$ and $B$ with respective temperatures $T_A$ and $T_B$. As an example of a system of this type, consider bacteriorhodopsin, a light-harvesting molecular machine found in certain microorganisms which has recently been analyzed through the lens of stochastic thermodynamics~\cite{pinero2024optimization}, and is typically modelled using a set of discrete states, with only a single transition mediated by light~\cite{lorenz2009spectroscopic}.

The entropy production rate can be decomposed into non-negative contributions due to the dynamics respectively coupled to each of the two reservoirs (Appendix A provides proof):
\begin{subequations}
\begin{align}
\dot{\Sigma} & = \dot{\Sigma}_A + \dot{\Sigma}_B,\\
& = \underbrace{\dot{S}_A - \frac{\dot{Q}_A}{\kB T_A}}_{\geq 0} + \underbrace{\dot{S}_B - \frac{\dot{Q}_B}{\kB T_B}}_{\geq0}.
\end{align}
\end{subequations}
Here $\dot{S}_A$ ($\dot{S}_B$) is the rate of change of the system entropy $S[Z]$ due to transitions coupled to the $A$ ($B$) reservoir, and $\dot{Q}_A$ ($\dot{Q}_B$) is the rate of heat from the $A$ ($B$) reservoir into the system. At steady state the system entropy is constant, so $0 = \mathrm{d}_t S = \dot{S}_A + \dot{S}_B$. Combining this with the reservoir-specific second laws and the steady-state first law $-\dot{W} = \dot{Q}_A + \dot{Q}_B$, we derive an IFAR-like result:
\begin{equation}
-\dot{W} \leq \kB(T_A - T_B)\dot{S}_A.
\end{equation}
We call this generalized result the \emph{entropy arbitrage relation} (EAR), which holds for nonequilibrium systems in contact with two thermal reservoirs, without requiring the bipartite structure. When the system is bipartite, such that it can be decomposed into two subsystems each in contact with a unique reservoir (as detailed in Sec.~\ref{sec:bipartite_stoch_td}), then the entropy rate $\dot{S}_A$ is equivalently the information flow from one subsystem to the other. Thus the IFAR emerges as the bipartite specialization of the EAR.

This inequality can be recast in terms of an information rate by considering the self-information $I[Z;Z] = S[Z]$~\cite[Sec.~2.4]{Cover2006_Elements}, so that
\begin{equation}\label{eq:gear}
-\dot{W} \leq \kB(T_A - T_B)\dot{I}_A.
\end{equation}
The information rate $\dot{I}_A$ is the rate at which the transitions coupled to the $A$ reservoir change the self-information $I[Z;Z]$. We call this quantity an information rate (rather than a flow) because it cannot always be considered a flow of information from one part of the system to another.

\section{Model systems and applications}\label{sec:models}

Using the theory of bipartite stochastic thermodynamics outlined above, we now analyze explicit nonequilibrium models to illustrate our main results: we consider the Brownian-gyrator heat engine~\cite{filliger2007brownian} modified to incorporate nonconservative driving forces, and a double quantum dot which constitutes a simple model for an autonomous information engine. These models are analytically tractable, allowing us to show explicitly that the Brownian gyrator must use information flow to achieve net output power, and the quantum-dot information engine must act as a heat engine to deliver positive output power. Finally, we explore applications of our results as tools for thermodynamic inference: our theory predicts significant information flow within light-harvesting molecular machines like photosystem II and bacteriorhodopsin, which we find is supported by experimentally parameterized stochastic models of their reaction dynamics.

\subsection{Brownian-gyrator heat engine} \label{sec:Brownian_gyrator}
\begin{figure*}[tb]
    \centering
    \includegraphics[width = \textwidth]{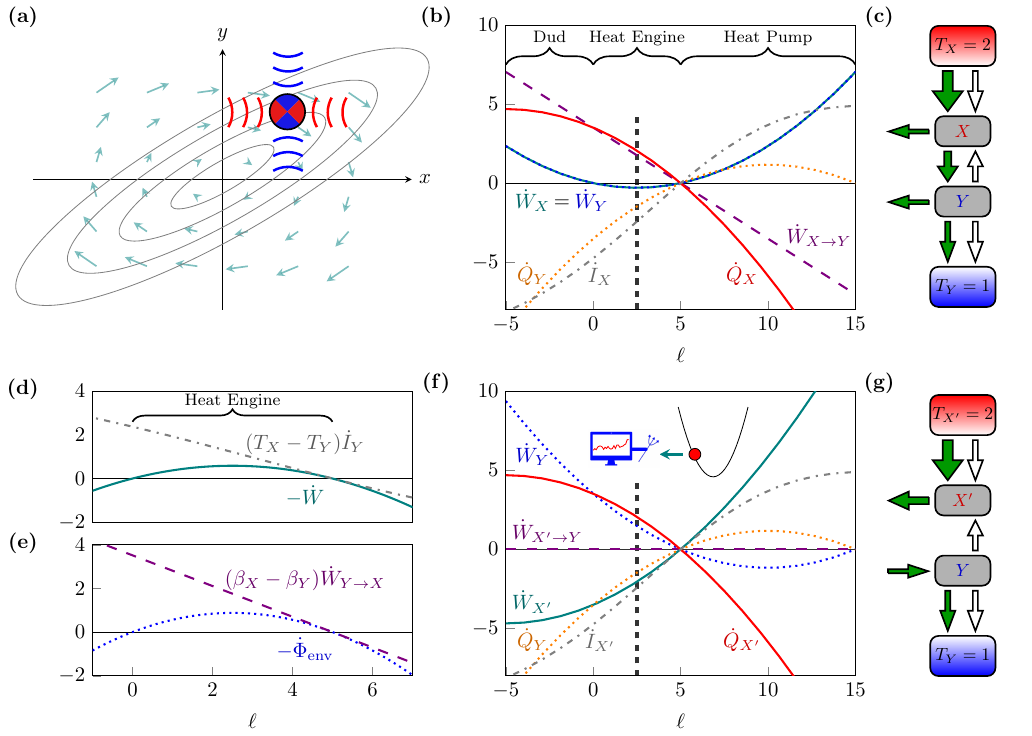}
    \caption{Steady state dynamics and thermodynamics of the Brownian gyrator. (a) Schematic diagram. Gray closed curves denote isopotential contours; light green arrows show the nonconservative force field; red-blue circle denotes the 2D Brownian particle; blue and red arcs denote different-strength fluctuations in different directions. 
    (b) Energy and information flows.
    (c) Thermodynamic diagram showing directionality of energy, entropy, and information flows for $\ell=2.5$ (black dashed vertical line in (b)). 
    (d,e) Verification of the two arbitrage relations, IFAR~\eqref{eq:IFAR} (d) and TPAR~\eqref{eq:TPAR} (e). 
    (f) Energy and information flows in the feedback-cooling information-engine interpretation. Inset: schematic, with nonconservative force from controller indicated by green arrow.
    (g) Thermodynamic diagram showing directionality of energy, entropy, and information flows for $\ell=2.5$ (black dashed vertical line in (f)).
    Throughout, $k=15$, $\tau=1/2$, and $\nu=1$.} 
    \label{fig:BrownianGyrator}
\end{figure*}

Consider the Brownian gyrator, a microscopic, stochastic model for a  steady-state heat engine depicted in Fig.~\ref{fig:BrownianGyrator}(a). First introduced in a slightly different form by Filliger and Reimann~\cite{filliger2007brownian}, the Brownian gyrator has since been studied extensively both in its original formulation~\cite{allahverdyan2009thermodynamic,dotsenko2013two,baldassarri2020engineered,miangolarra2021energy,miangolarra2023minimal} along with a plethora of different extensions including the addition of conservative~\cite{cerasoli2018asymmetry,fogedby2017minimal} or nonconservative~\cite{pietzonka2018universal,lin2022stochastic} external forces, higher-order potentials~\cite{chang2021autonomous}, underdamped dynamics~\cite{bae2021inertial}, as well as non-Markovian~\cite{dos2021stationary} and active~\cite{lee2020brownian} fluctuations. The dynamics of the Brownian gyrator can also be mapped directly onto electric-circuit models with resistors subject to Johnson noise from different heat baths; this equivalent system and several extensions have been thoroughly studied both theoretically and experimentally~\cite{ciliberto2013heat,ciliberto2013statistical,chiang2017electrical,miangolarra2022thermodynamic,miangolarra2023matching}. The Brownian gyrator has also been realized experimentally as an overdamped particle with electromagentically induced anisotropic fluctuations~\cite{argun2017experimental,abdoli2022tunable}.

Our formulation of the Brownian gyrator, illustrated in Fig.~\ref{fig:BrownianGyrator}(a), is equivalent to that presented in Ref.~\cite{pietzonka2018universal}, a bipartite system with two degrees of freedom $X$ and $Y$ whose dynamics evolve according to the coupled overdamped Langevin equations
\begin{subequations}\label{bgeqs}
\begin{align}
\dot{x} & = \ell y - \partial_xV(x,y) + \sqrt{2}\,\eta_X(t),\\
\dot{y} & = \nu\left[-\ell x - \partial_y V(x,y)\right] + \sqrt{2\nu\tau}\,\eta_Y(t).
\end{align}
\end{subequations}
Here $\eta_X(t)$ and $\eta_Y(t)$ are uncorrelated Gaussian white noise sources with $\langle \eta_X(t) \eta_X(t') \rangle = \delta(t-t')$ and similarly for $\eta_Y(t)$, $\nu$ is the ratio of the two mobility coefficients, $\tau=T_Y/T_X$ is the temperature ratio, and the potential is
\begin{equation}
V(x,y) = \frac{1}{2}x^2 + \frac{1}{2}y^2+\frac{1}{2}k(x-y)^2,
\end{equation}
for coupling strength $k$ between $X$ and $Y$. Furthermore, $f_X(x,y) = \ell y$ and $f_Y(x,y) = -\ell x$ are nonconservative forces of strength $\ell$ that, on their own, induce a rotation of the system in the $x\mathrm{-}y$ plane. All quantities are dimensionless. The nonconservative forces were not present in the original formulation of the Brownian gyrator~\cite{filliger2007brownian}, and have been added here as in Ref.~\cite{pietzonka2018universal} so that work can be input to or extracted from the Brownian gyrator in a thermodynamically consistent manner. Such nonconservative forces can be incorporated into electrical implementations of the gyrator through, for example, a non-reciprocal capacitor~\cite{miangolarra2023matching}.

The coupled Langevin equations~\eqref{bgeqs} are linear, so the stationary joint probability distribution for $X$ and $Y$ can be solved analytically~\cite{risken1996fokker}, from which the ensemble-averaged energy and information flows can be computed analytically from their definitions~\cite{leighton2023inferring} using computer algebra software (plotted in Fig.~\ref{fig:BrownianGyrator}).

Figure~\ref{fig:BrownianGyrator}(b) illustrates the energy and information flows in the Brownian gyrator. In the regime of heat-engine operation, when $0<\ell<k(1-\tau)/(1+\tau)$, both subsystems ($X$ and $Y$) output work at positive rates ($\Wx<0$ and $\Wy<0$). This is powered by a flow of heat into the hotter $X$ subsystem ($\Qx>0$), which by the second law then requires a net flow of heat out of the cooler $Y$ subsystem ($\Qy<0$). As required by the arbitrage relations~\eqref{eq:IFAR} and~\eqref{eq:TPAR}, the information flow and transduced power are both non-zero, with information flowing from cold to hot ($\Ix<0$) and transduced work flowing from hot to cold ($\Wxy>0$). This thermodynamic setup is shown in Fig.~\ref{fig:BrownianGyrator}(c).

For $k(1-\tau)/(1+\tau)<\ell<k$, the Brownian gyrator operates as a heat pump, with net input work into both subsystems ($\Wx>0$ and $\Wy>0$) powering the flow of heat from cold to hot ($\Qy>0$ and $\Qx<0$). Here the information flow goes from hot to cold ($\Ix>0$), a condition which follows directly from the second law applied to bipartite heat pumps. For $-k<\ell<0$ the Brownian gyrator is a dud, with input work into both subsystems accompanying a flow of heat from hot to cold. Finally for $|\ell|>k$, the setup is again a dud (albeit of a different sort), with input work into both subsystems resulting in heat flows into both reservoirs.

Figures~\ref{fig:BrownianGyrator}(d) and (e) explicitly confirm the two arbitrage relations in this system. We find that the inequalities are tighter when the system is near equilibrium. For the Brownian gyrator this is at stall, where all energy and information flows vanish, for $\ell=k(1-\tau)/(1+\tau)$.

If, as we claim, information flow along with a temperature difference is what drives the net power output in the Brownian gyrator, we should be able to find the information engine hidden in this setup. To do this, we rescale the variable $X$ such that $x'\equiv x(k-\ell)/(1+k)$, and define the parameters $\kappa\equiv 1+k$, $\tau_m\equiv 1/[\nu(1+k)]$, $\sigma^2 \equiv  2\tau/[\nu(1+k)^2]$, and $a\equiv (\ell^2-k^2)/(1+k)$.

The resulting dynamics are equivalent to the feedback-cooling information-engine model studied by Horowitz and Sandberg~\cite{horowitz2014second}, with the controlled system consisting of the position $x'$ of an overdamped Brownian particle in a quadratic trap with strength $\kappa$ and the controller $y$ monitoring the dynamics of the system:
\begin{subequations}\label{infoengeqs}
\begin{align}
\dot{x}' & = -\kappa x' - a y + \xi_{X'},\\
\tau_m\dot{y} & = -y+x' + \xi_Y,
\end{align}
\end{subequations}
The noise terms $\xi_X$ and $\xi_Y$ correspond to independent Gaussian white noise with respective variances $2 T_X$ and $\sigma^2$, $a$ is the feedback gain, and $\tau_m$ is a time constant by which $Y$ can be considered to low-pass filter noisy measurements of $X$ with measurement noise $\sigma^2$. 

In addition to rescaling the $X$ variable and redefining the various parameters, we also adjust our interpretation of the sources of the forces acting on the two subsystems. In keeping with the interpretation of Ref.~\cite{horowitz2014second}, we take $V(x',y)=\frac{1}{2}\kappa (x')^2$ to be the conservative potential, and $f_X(x',y)=-ay$ and $f_Y(x',y)=x'-y$ to be the nonconservative forces. This change in perspective from the original Brownian-gyrator interpretation can be thought of as a gauge transformation, as considered in Ref.~\cite{ding2022unified}. The information flow is unchanged by rescaling one of the variables (because the mutual information itself is invariant under variable rescaling~\cite{kraskov2004estimating}), so we identify the information flow within the Brownian gyrator as the same information flow found within the feedback-cooling information engine.

Figure~\ref{fig:BrownianGyrator}(f) shows the thermodynamics of the feedback-cooling information engine, as quantified by the energy and information flows. The heat and information flows are unchanged from those of the Brownian gyrator, but the input, output, and transduced powers are modified. In particular, since there is no potential energy coupling the controller to the particle, the transduced power $\Wxy$ is always zero. For $\ell<k(1-\tau)/(1+\tau)$, output work is extracted from the particle ($\Wx<0$), while input work is required to run the controller ($\Wy>0$).

\subsection{Double-quantum-dot information engine} \label{sec:quantum_dot}
Our treatment of the Brownian gyrator focused on interpreting a bipartite heat engine as an information engine. Here we illustrate the converse, showing that an established model of an autonomous information engine can only deliver positive total output power if the controller and feedback-controlled system are at different temperatures, thus rendering the setup a bipartite heat engine.

\begin{figure*}[tb]
    \centering
    \includegraphics[width = \textwidth]{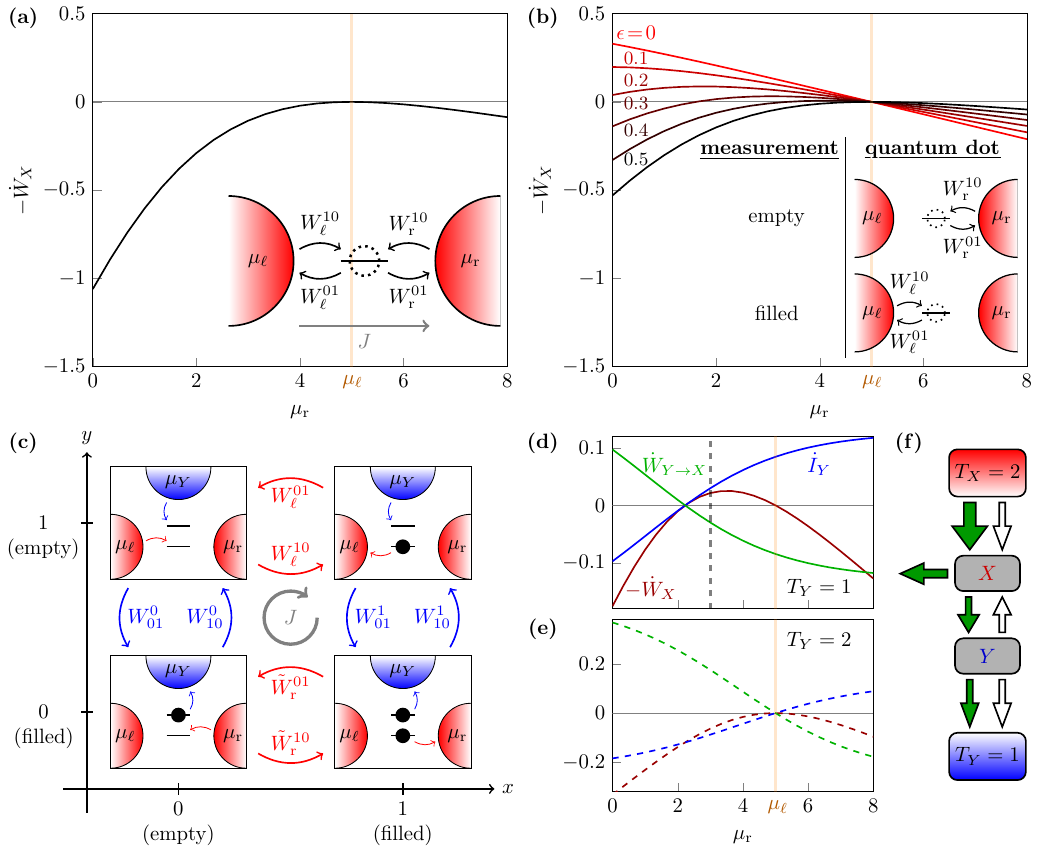}
    \caption{Steady state thermodynamics of the quantum-dot information engine. 
    (a) Output power $-\dot W_X$ by a single quantum dot in simultaneous contact with two reservoirs at temperature $T_X=2$ and respective chemical potentials $\mul=5$ and $\mur$, as a function of $\mur$, for $\Gamma_X^\ell = \Gamma_X^\mathrm{r}=1$. Inset: transition rates for electrons hopping into and out of the quantum dot.
    (b) When measurement with error $\epsilon$ and feedback are added, $-\dot W_X>0$ is possible, indicating that electrons are pumped against their natural gradient.
    (c) State diagram and possible transitions for the double-quantum-dot model.
    (d) Output power $-\dot W_X$, transduced power $\Wyx$, and information flow $\dot I_Y$ of in the double quantum dot for the same parameters as in (a) and (b) and $\epsilon=0.2$, $T_Y=1$, $\Gamma_Y=10^3$. (e) Same plot but for $T_Y=2$.
    (f) Thermodynamic diagram showing directionalities of energy, entropy, and information flows for $\mur=3$ and $T_Y=1$ [gray dashed vertical line in (d)].
    Throughout: faint vertical orange line indicates $\mu_\ell$.} 
    \label{fig:QuantumDot}
\end{figure*}

Consider a single quantum dot $X$ in contact with two reservoirs (leads) at temperature $T_X$ and with chemical potentials $\mul$ for the left lead and $\mur$ for the right lead [Fig.~\ref{fig:QuantumDot}(a) inset]. Electrons can jump between either reservoir and the quantum dot, which can be either empty ($x=0$) or filled ($x=1$). The rates satisfy detailed balance with the respective reservoir:
\begin{subequations}
\begin{align}
W^{10}_\ell &= \frac{\Gamma_X^\ell}{e^{-\beta_X\mul}+1}\,, \quad W^{01}_\ell = \frac{\Gamma_X^\ell\,e^{-\beta_X\mul}}{e^{-\beta_X\mul}+1} \label{eq:transition_rates_L}\\
W^{10}_\mathrm{r} &= \frac{\Gamma_X^\mathrm{r}}{e^{-\beta_X\mur}+1}\,, \quad W^{01}_\mathrm{r} = \frac{\Gamma_X^\mathrm{r}\,e^{-\beta_X\mur}}{e^{-\beta_X\mur}+1} \label{eq:transition_rates_R}
\end{align}
\end{subequations}
where $\Gamma_X^\ell/2$ and $\Gamma_X^\mathrm{r}/2$ are bare rate constants, i.e., the rate constants at equilibrium (when $\mul=0=\mur$). Here and throughout, the superscript ``$10$" denotes the transition from state $0$ to state $1$, with ``$01$" denoting the reverse transition. Note that in this subsection the symbol $W$ with a subscript and a superscript denotes a transition rate, not to be confused with $\dot{W}$ with a subscript that throughout this paper denotes power.

The average current of electrons flowing from the left lead is $J_{\rm \ell} \equiv W^{10}_\ell p_0-W^{01}_\ell p_1$, where $p_0$ and $p_1$ are the respective stationary probabilities for the dot being empty and filled. Similarly, $J_{\rm r} \equiv W^{01}_\mathrm{r}p_1-W^{10}_\mathrm{r}p_0$. Solving the system of equations consisting of the definitions of $J_{\ell}$ and $J_{\rm r}$, the steady-state equality $J_{\ell}=J_{\rm r}=J$, and  normalization $p_0+p_1=1$, yields $p_0$, $p_1$, and $J$ as functions of the chemical potentials $\mul$ and $\mur$, temperature $T_X$, and bare rate constants $\Gamma_X^\mathrm{r}$ and $\Gamma_X^\ell$. These then allow us to calculate the steady-state power from the quantum dot to the reservoirs,
\begin{align}
    -\Wx \equiv J\,\left(\mur-\mul\right)\,.
    \label{eq:QD_power}
\end{align}
This power is equal to the heat flow from the reservoir due to the conduction of electrons through the quantum dot. Figure~\ref{fig:QuantumDot}(a) depicts this power as a function of the chemical potential $\mur$ for fixed $\mul$, $T_X$, and $\Gamma_X^{\ell, \mathrm{r}}$.

``Maxwell-demon feedback''~\cite{Schaller2012_Stochastic} can be used to pump electrons from right to left, against the chemical-potential gradient. Like Maxwell's original demon~\cite{Maxwell1867} that implements feedback by opening and closing a trap door between two gas volumes, this feedback only modifies the bare transition rates, not the energy levels. We assume that the state of the quantum dot is continuously measured by an auxiliary measurement device that modifies the bare rates $\Gamma_X$ such that $\Gamma_X^\ell=0$ when the dot is measured empty and $\Gamma_X^\mathrm{r}=0$ when the dot is measured filled [Fig.~\ref{fig:QuantumDot}(b) inset].

The measurement has an error probability $\epsilon$ with which it erroneously measures the opposite state. Then the net flux from left to right is 
\begin{subequations}
\begin{align}
J & =W^{10}_\ell p_0 \epsilon-W^{01}_\ell p_1 (1-\epsilon)\\
& = W^{01}_\mathrm{r} p_1 \epsilon- W^{10}_\mathrm{r} p_0 (1-\epsilon).
\end{align}
\end{subequations}
Solving this equation with normalization $p_0+p_1=1$ yields the steady-state occupation probabilities and the flux as functions of $\mul$, $\mur$, $T_X$, and $\epsilon$. Equation~\eqref{eq:QD_power} then gives the steady-state output power $-\Wx$ [Fig.~\ref{fig:QuantumDot}(b)]. For $\epsilon<1/2$, this power can become positive, indicating that the quantum dot delivers power to the reservoirs. This output power stems from the heat due to thermal fluctuations that spontaneously fill the dot with an electron from the right reservoir. The feedback then rectifies these fluctuations by preferentially allowing the electron to flow into the left reservoir.

The quantum dot and the feedback mechanism together constitute a Maxwell-demon setup or information engine. However, we have yet to specify the physical mechanism of the controller. As pointed out in previous works~\cite{strasberg2013thermodynamics,horowitz2014thermodynamics,Kutvonen2016_Thermodynamics,Sanchez2019_Autonomous,Annby-Andersson2020_Maxwells,Tanogami2023_Universal}, a second capacitively coupled quantum dot can serve as the controller for this information engine. Consider the single-level quantum dot $Y$ that is coupled to a reservoir with chemical potential $\mu_Y$ and temperature $T_Y$. Somewhat counterintuitively, a filled $Y$ ($y=0$) can encode the measurement of an empty $X$ ($x=0$) and an empty $Y$ ($y=1$) can encode the measurement of a filled $X$ ($x=1$). The capacitive interaction between the dots results in a potential energy $U$ when both dots are filled. Otherwise, the potential energy vanishes. 

Figure~\ref{fig:QuantumDot}(c) depicts the transition rates between the four different configurations of the double quantum dot. The transition rates into and out of the $X$-dot are 
\begin{align}
    W^{xx'}_y = \begin{cases}
        \tilde W^{xx'}_\mathrm{r} \ ,& y=0 \text{ (filled)}\\
        W^{xx'}_\ell \ ,& y = 1 \text{ (empty)}\,,
    \end{cases} \label{eq:transition_rates_X^DQD}
\end{align}
where $x$ and $x'$ are the states of $X$ with $x\neq x'$, $W^{xx'}_\ell$ is given by Eq.~\eqref{eq:transition_rates_L}, and
\begin{subequations}\label{eq:QD_modified_R_rates}
\begin{align}
    \tilde W^{10}_\mathrm{r} &= \frac{\Gamma_X^\mathrm{r}}{e^{-\beta_X(\mur-U)}+1}\,, \;\\
    \tilde W^{01}_\mathrm{r} & = \frac{\Gamma_X^\mathrm{r}\,e^{-\beta_X(\mur-U)}}{e^{-\beta_X(\mur-U)}+1}
\end{align}
\end{subequations}
correspond to the rates in Eq.~\eqref{eq:transition_rates_R}, modified by the potential energy $U$ due to the interaction with the other quantum dot $Y$. Equation~\eqref{eq:transition_rates_X^DQD} implies that the $X$-dot couples to different reservoirs depending on the state of the $Y$-dot. This is a special case of the treatment in, e.g., Refs.~\cite{strasberg2013thermodynamics,horowitz2014thermodynamics} in which the bare rates $\Gamma_X^\ell$ and $\Gamma_X^\mathrm{r}$ are modified through a $Y$-dependent density of states. Our model thus corresponds to an idealized double-quantum-dot information engine with only one global cycle and no local cycles~\cite{horowitz2014thermodynamics}.

To implement the measurement, the $Y$-transitions are governed by
\begin{subequations}
\begin{align}
    W^0_{01} &= \frac{\Gamma_Y}{e^{-\beta_Y\mu_Y}+1}\,, \quad\;\;\;\; W^0_{10} = \frac{\Gamma_Y\,e^{-\beta_Y\mu_Y}}{e^{-\beta_Y\mu_Y}+1} \\
    W^1_{01} &= \frac{\Gamma_Y}{e^{-\beta_Y(\mu_Y-U)}+1}\,, \; W^1_{10} = \frac{\Gamma_Y\,e^{-\beta_Y(\mu_Y-U)}}{e^{-\beta_Y(\mu_Y-U)}+1}\,
\end{align}
\end{subequations}
where $U=2\mu_Y$ and $\mu_Y = \kB T_Y\ln\left[(1-\epsilon)/\epsilon\right]$ are chosen such that
\begin{align}
    \frac{W^{0}_{01}}{W^0_{10}} = \frac{W^{1}_{10}}{W^1_{01}} =\frac{1-\epsilon}{\epsilon},
\end{align}
and hence the measurement-error probability is $\epsilon$. Setting $\Gamma_Y \gg \Gamma_X^{\mathrm{r},\ell}$ ensures that $Y$ quickly relaxes into a local equilibrium distribution corresponding to the desired measurement distribution.

The net current of electrons from the left to right leads is determined by solving
\begin{align}
    J &= W^{10}_\ell p_{01} - W^{01}_\ell p_{11} = W^{1}_{01} p_{11} - W^{1}_{10} p_{10} \nonumber\\
    &= \tilde W^{01}_\mathrm{r} p_{10} - \tilde W^{10}_\mathrm{r} p_{00} = W^{0}_{10} p_{00} - W^{0}_{01} p_{01}
\end{align}
along with normalization $p_{00} + p_{10} + p_{10} + p_{11}=1$ constraining the probability $p_{xy}$ to find the joint system in state $(x,y)$. Figure~\ref{fig:QuantumDot}(d) shows the net power~\eqref{eq:QD_power} done on $X$ by the two $X$-reservoirs. 

Importantly, the output power $-\Wx$ differs from the case where the controller does not require power [the $\epsilon = 0.2$ curve in Fig.~\ref{fig:QuantumDot}(b)]. Specifically, positive output power ($-\Wx>0$) is only possible for $2 \lesssim \mur \lesssim 5$. This is because power is required to run the controller $Y$, which itself is not directly driven by a chemical potential difference since it has access to only one reservoir, and thus $\dot W_Y = J\, (\mu_Y - \mu_Y) = 0$. As depicted in Fig.~\ref{fig:QuantumDot}(f), the power to run the controller $Y$ is diverted from the output power as transduced power $-\Wyx$ [Fig.~\ref{fig:QuantumDot}(d), green curve]. The blue curve in Fig.~\ref{fig:QuantumDot}(d) shows the information flow $\dot I_Y$, with which we can verify the IFAR~\eqref{eq:IFAR},
\begin{align}
    -\dot W_X - \underbrace{\dot W_Y}_{=0} \leq \underbrace{\left(T_X - T_Y\right)}_{=1}\, \dot I_Y\,. \label{eq:IFARDot}
\end{align}

Finally, Fig.~\ref{fig:QuantumDot}(e) shows output power $-\dot W_X$ and transduced power $\Wyx$ when both quantum dots are at equal temperature ($T_Y=T_X=2$), so the RHS of \eqref{eq:IFARDot} vanishes and $-\dot{W}_X \leq 0$. In this case so much power $-\Wyx$ is diverted that no positive output power can be generated. This illustrates that the double-quantum-dot information engine can only deliver positive output power when the controller is at a lower temperature than the feedback-controlled system, i.e., when the joint system operates as a bipartite heat engine, as predicted by the information-flow arbitrage relation~\eqref{eq:IFAR}.

\subsection{Inferring information flows in light-harvesting molecular machines}
\label{PSIIInference}
In addition to elucidating the duality of heat engines and information engines, the arbitrage relations introduced in this article can also be used for thermodynamic inference~\cite{seifert2019stochastic}. Using the IFAR~\eqref{eq:IFAR}, observing net output power ($\dot{W}<0$) from a bipartite system immediately implies the existence of both a temperature difference and an information flow, whose sign is further implied if the ordering of the two temperatures is also known. Through TPAR~\eqref{eq:TPAR}, the existence and directionality of internally transduced power can likewise be inferred in autonomous systems at steady state. More quantitatively, each of the IFAR, TPAR, and EAR~\eqref{eq:gear} can be rearranged to yield a bound on internal flows of energy or information, requiring only knowledge of (often experimentally tractable) input and output works and the two reservoir temperatures. For example, IFAR can be rearranged to yield a lower bound on the information flow inside a bipartite heat engine:
\begin{equation}
\dot{I}_Y \geq \frac{-\dot{W}}{\kB(T_X-T_Y)}.
\end{equation}

We illustrate this application by inferring the existence and magnitude of the information rate inside photosystem II, one of the molecular machines responsible for photosynthesis in plant cells. Photosystem II is in contact with the ambient cellular environment ($T_Y\approx 300$K) as well as hot solar radiation emitted from the surface of the sun at $\approx 5800$K~\cite{dorfman2013photosynthetic}. Its dynamics include light-induced electronic transitions of the P680 complex (coupled to a high-temperature thermal reservoir at $T_X\leq 5800$K) and water-splitting chemical reactions of the oxygen-evolving complex (OEC)~\cite{vinyard2013photosystem} coupled to the lower-temperature thermal reservoir at $T_Y$. While models of photosystem-II dynamics differ on whether the dynamics of P680 and OEC satisfy the bipartite assumption~\cite{lazar2003chlorophyll,jablonsky2008evidence,lazar2009approaches}, they uniformly ascribe each transition to a particular reservoir. Estimating the mean output work from the free-energy change $\approx$\,237\,kJ/mol~\cite{ye2019artificial} and net reaction rate $\approx$\,350\,/s~\cite{forbush1971cooperation}, we infer a minimum information rate of $\approx$\,7\,bit/reaction (or equivalently $\approx$\,2000\,bit/s) inside photosystem II. In general this is an information rate due to the dynamics of light-induced transitions, however under the more restrictive bipartite assumptions of IFAR this is more specifically an information flow between the OEC and P680 (the rate is the same, only the interpretation differs).

While available experimental data for photosystem II is insufficient to directly quantify information flows due to photochemical dynamics, detailed stochastic models for the reaction dynamics have been constructed and fit to experimental data. A popular such model, Lazár and Jablonsky's Scheme 4~\cite{lazar2009approaches}, incorporates both photophysical dynamics of the P680 complex and chemical dynamics of the OEC. This model is not bipartite, but uniquely identifies transitions as coupled to either photon absorption/emission or chemical reaction dynamics. Note that this model violates our assumption of thermodynamic consistency: several model transitions are irreversible. In this experimentally parameterized stochastic model for photosystem II dynamics, we calculate an information rate of $\approx$\,9\,bit/reaction (Appendix B provides calculational details), above but remarkably close to our model-agnostic lower bound.

In addition to photosystem II our results apply to other light-harvesting molecular machines like bacteriorhodopsin, which uses free energy from sunlight to pump protons across membranes in diverse species of archaea. The reaction dynamics and thermodynamics of bacteriorhodopsin are well understood~\cite{varo1991thermodynamics,lorenz2009spectroscopic}, and while the reaction cycle is not bipartite, models nonetheless uniquely couple transitions to either solar photons ($T_X\leq 5800$K as in photosystem II) or the ambient cellular environment ($T_Y\approx 293$K). Using a typical output work rate of $\approx$\,6.1\,$\kB T_\mathrm{cell}$/cycle~\cite{pinero2024optimization}, the EAR predicts an information rate with magnitude $\gtrsim$\,0.5\,bit/cycle. Solving the master equation for the model used in Ref.~\cite{pinero2024optimization}, we compute an information rate of $2$\,bit/cycle (Appendix B provides calculational details), in agreement with our model-agnostic prediction and comparable but somewhat lower than that found in photosystem II.

\section{Discussion} \label{sec:discussion}
This paper illustrates that functioning bipartite heat engines must transmit information (i.e., a reduction in entropy) between subsystems in contact with heat reservoirs at different temperatures in order to produce net output work, through a process analogous to economic arbitrage. This implies that they are also information engines, in the sense that they sustain an information flow that powers an apparent violation of the second law. This implies that the field of information thermodynamics~\cite{Parrondo2015_Thermodynamics} applies to real-world heat engines.

Our findings directly imply design principles for nanoscale systems, like molecular machines, operating in environments with inhomogeneous or anisotropic fluctuations. When these systems are composed of different parts each in contact with different strengths of fluctuations, maximizing output power requires these components to operate collectively and exchange entropy in the form of information flows. This leads to ``Maxwell-demon" behavior, where one component extracts heat from its environment in apparent, but not true, violation of the second law. Thus Maxwell's demon may well lie hidden within biological molecular machines which have evolved to take advantage of different sources of fluctuations in the cellular environment.

Conversely, this paper also illustrates that information engines can deliver positive net output power when controller and controlled system are at different temperatures. This fact was explored in Refs.~\cite{Still2020_cost_benefit_memory,Still2021_Partially}, where the ratio of the temperatures of the work medium and memory parameterizes optimal information-processing strategies in variants of the Szilard engine. It is also implicit in the analysis of Ref.~\cite{Saha2022_Information}, where an information engine delivers net power derived from active fluctuations that mimic the effect of a larger temperature. 

Our framework helps to demystify information engines by providing a change in perspective that illustrates they are variants of heat engines, in which the entropy reduction step is ``outsourced'' to an auxiliary controller or memory. Ignoring this auxiliary system leads to an apparent second-law violation, highlighting the importance of accounting for the thermodynamic costs of the controller's entropy reduction. The joint setup of controller and controlled system can only deliver positive net output work when the controller is at a lower temperature, thereby giving a heat engine that delivers net power by exchanging heat with two reservoirs at different temperatures. While previous theoretical analyses have hinted at this connection for specific systems~\cite{feynman1965feynman,parrondo1996criticism,Still2020_cost_benefit_memory,Still2021_Partially}, our results here are far broader, encompassing fully general mathematical proofs and intuitive explanations.

In addition to proving our main results using the theory of bipartite stochastic thermodynamics, we also illustrated them intuitively using an analogy to economics -- providing a qualitative argument accessible without reference to stochastic thermodynamics. The concept of arbitrage lends itself well to understanding both classical and bipartite heat engines: the temperature of a heat bath can be thought of as the ``exchange rate" between energy and entropy. By ``trading" energy and entropy with different baths (``markets"), a heat engine can perform ``arbitrage" to produce net output work (``profit"). Such a scheme requires moving energy and entropy from one bath to another, leading to the requirements for information flows and transduced power respectively quantified by the IFAR~\eqref{eq:IFAR} and TPAR~\eqref{eq:TPAR}. The usefulness of this analogy should not be surprising; after all thermodynamics is fundamentally the science of accounting for energy and entropy. Other analogies have likewise been drawn between stochastic thermodynamics and economics~\cite{smerlak2016thermodynamics,ducuara2023maxwell}; exploring such analogies further, for example at a more quantitative level, could lead to deeper insights into both fields.

As shown in Sec.~\ref{environment_potential_interpretations}, our main results [the IFAR~\eqref{eq:IFAR} and TPAR~\eqref{eq:TPAR}] can be reinterpreted as constraining changes in thermodynamic potentials of the environment encompassing the two thermal reservoirs. Since the reservoirs by assumption are at equilibrium, and do not directly support correlations or interactions with each other, it is straightforward to define their free energies (Helmholtz potentials) and free entropies (Massieu potentials) even when doing so for the system itself is not possible due to its nonequilibrium state and contact with multiple temperatures. These thermodynamic potentials are particularly useful, allowing for a qualitative understanding of our main results without the more involved theoretical machinery of stochastic thermodynamics; such an approach may prove useful for considering other nonequilibrium systems for which thermodynamic potentials cannot easily be defined.

An interesting future challenge would be to relax the bipartite assumption and derive the analogs of IFAR~\eqref{eq:IFAR} and TPAR~\eqref{eq:TPAR} from the information-flow formalism for systems without bipartite structure~\cite{chetrite2019information}. Likewise, expanding our work to cover information reservoirs~\cite{Deffner2013_Information,Barato2014_Information_Reservoirs} and the many models of information engines that are based on them (e.g.,~\cite{Mandal2012_solvable_model,Horowitz2013_Imitating,Mandal2013_Refrigerator}) should be possible. Future work could also explore the information-theoretic requirements for leveraging correlated (athermal) noise sources~\cite{pietzonka2019autonomous}.

Beyond elucidating design principles, the arbitrage relations (IFAR, TPAR, and EAR) are also powerful tools for thermodynamic inference~\cite{seifert2019stochastic}. In particular, using only information about temperatures and external work rates, these arbitrage relations can be used to infer the existence and magnitude of internal energy and information flows within molecular machines. We illustrated this potential for model-agnostic inference in Sec.~\ref{PSIIInference} by estimating the magnitude of information flow in the molecular machine photosystem II, which we argue can be considered a bipartite heat engine coupled to both photons at the high temperature of the solar blackbody spectrum and chemical reactions at the much cooler ambient cellular temperature. Our prediction, a lower bound on the information flow of $\approx$\,7\,bits per reaction cycle ($\approx$\,2000\,bit/s), is validated by computation of the information rate ($\approx$\,9\,bits per reaction cycle) in an experimentally parameterized stochastic model of the photosystem II reaction cycle~\cite{lazar2009approaches}. We similarly apply our results to infer a significant information rate ($\approx$\,5\,bit/s) in another light-harvesting molecular machine, bacteriorhodopsin, which we likewise verify through computational modelling. The magnitude of information rates found in these light-harvesting molecular machines is striking when compared to other biological information rates like that underlying bacterial chemotaxis, estimated at 0.03\, bit/s~\cite{mattingly2021escherichia}, and those found in biochemical signalling networks, on the order of bits per hour~\cite{cheong2011information}. It would be interesting to explore systematic variation of this information rate across different classes of molecular machines.

Finally, we step back to consider macroscopic heat engines in simultaneous contact with two heat baths at different temperatures, for example thermoelectric devices. Since none of our main theoretical results in Sec.~\ref{sec:bipartite_stoch_td} are built on assumptions about system size, our conclusions should still hold for macroscopic systems. This then raises the obvious question of where the information flow predicted by the IFAR can be found in, for example, a thermoelectric generator. We conjecture that the information flow is encoded in the statistics of electron positions and momenta~\cite{freitas2023information}. A first step could be to consider small-scale systems which allow exact counting of electrons in a thermoelectric device. For larger thermoelectric devices, the electron positions and momenta are aggregated into correlations of voltage and current fluctuations across the two junctions in contact with different temperatures. Measurement of voltage and current fluctuations would be analogous to measuring the pressure in each quadrant of the two-dimensional ideal-gas Carnot engine in Sec.~\ref{sec:2dcarnotgas}, which aggregates the statistics of the $N$ gas molecules. The information flow in the engine could be extracted from pressure measurements in each quadrant with fine temporal resolution. It would be interesting to verify these predictions experimentally, namely attempting to measure current and voltage correlations in a thermoelectric device, and thus quantify the predicted information flow. Such a result would complement recent theoretical predictions of macroscopic information flows~\cite{freitas2022maxwell,freitas2023information}, and serve to illustrate that core concepts from stochastic thermodynamics such as information flows have real relevance in macroscopic systems, far beyond the nanoscale regime in which they were originally formulated.

\emph{Acknowledgments}.---
We thank Steven Blaber (UBC Physics) for helpful discussions, and John Bechhoefer, Eric Jones, and Johan du Buisson (SFU Physics), Eric Werker (SFU Beedie School of Business), and Oren Raz (Weizmann Physics) for feedback on the manuscript. This work was supported by a Natural Sciences and Engineering Research Council of Canada (NSERC) CGS Doctoral fellowship (M.P.L.), an NSERC Discovery Grant and Discovery Accelerator Supplement RGPIN-2020-04950 (D.A.S.), a Tier-II Canada Research Chair CRC-2020-00098 (D.A.S.), and grant FQXi-IAF19-02 from the Foundational Questions Institute Fund, a donor-advised fund of the Silicon Valley Community Foundation (J.E.\ and D.A.S.). M.P.L.\ thanks Howard and Caroline Malm for financial support awarded through the SFU Department of Physics.

\section*{Appendix A: Decomposing the second law without the bipartite structure}
Here we derive the second-law decomposition used in Sec.~\ref{section:gear_derivation} to derive the entropy arbitrage relation.

Consider a stochastic system $Z$ with a discrete set of states $\{z_i\}$, coupled to two distinct thermal reservoirs $A$ and $B$, whose dynamics follow the master equation
\begin{equation}
\frac{\partial}{\partial t} p_i = \sum_j \left[\left(R_{ij}^A + R_{ij}^B\right)p_j - \left(R_{ji}^A + R_{ji}^B\right)p_i\right].
\end{equation}
Here $R_{ij}^A$ and $R_{ij}^B$ denote transition rates respectively coupled to the two distinct thermal reservoirs $A$ and $B$ at respective temperatures $T_A$ and $T_B$.

The total entropy production rate, encompassing changes $\mathrm{d}_tS[Z]$ to the system Shannon entropy and heat flows $\dot{Q}_A$ and $\dot{Q}_B$ from the respective thermal reservoirs, is
\begin{subequations}
\begin{align}
\dot{\Sigma} & = \underbrace{\sum_{i>j}\left[ \left( R^A_{ij} + R^B_{ij}\right) p_j -\left( R^A_{ji} + R^B_{ji}\right)p_i \right] \ln\frac{p_j}{p_i}}_{\mathrm{d}_tS[Z]} \\
& + \underbrace{\sum_{i>j}\left[ R^A_{ij} p_j - R^A_{ji}p_i \right] \ln\frac{R^A_{ij}}{R^A_{ji}}}_{-\dot{Q}_A/\kB T_A} \\
& + \underbrace{\sum_{i>j}\left[ R^B_{ij} p_j - R^B_{ji}p_i \right] \ln\frac{R^B_{ij}}{R^B_{ji}}}_{-\dot{Q}_B/\kB T_B}.
\end{align}
\end{subequations}
This decomposition is similar to the analysis of Ref.~\cite{esposito2012stochastic}, which similarly identifies entropy production contributions due to different reservoirs. By splitting the system entropy change into contributions due to transitions respectively coupled to the $A$ and $B$ reservoirs,
\begin{equation}
\begin{aligned}
\mathrm{d}_tS[Z] & = \underbrace{\sum_{i>j}\left[R^A_{ij} p_j - R^A_{ji}p_i \right] \ln\frac{p_j}{p_i}}_{\dot{S}_A} \\
& + \underbrace{\sum_{i>j}\left[R^B_{ij} p_j - R^B_{ji}p_i \right] \ln\frac{p_j}{p_i}}_{\dot{S}_B},
\end{aligned}
\end{equation}
the total entropy production rate can be decomposed into respective contributions due to the $A$ and $B$ reservoirs:
\begin{equation}
\dot{\Sigma} = \underbrace{\dot{S}_A - \frac{\dot{Q}_A}{\kB T_A}}_{\dot{\Sigma}_A} + \underbrace{\dot{S}_B - \frac{\dot{Q}_B}{\kB T_B}}_{\dot{\Sigma}_B}.
\end{equation}

Like in the bipartite case, here $\dot{\Sigma}_A$ and $\dot{\Sigma}_B$ are both nonnegative:
\begin{subequations}
\begin{align}
\dot{\Sigma}_A & = \dot{S}_A - \frac{\dot{Q}_A}{\kB T_A}\\
& = \sum_{i>j}\left(R_{ij}^Ap_j-R_{ji}^Ap_i\right)\ln\frac{p_j}{p_i} \\
& \,\,+ \sum_{i>j}\left(R_{ij}^Ap_j-R_{ji}^Ap_i\right)\ln\frac{R_{ij}^A}{R_{ji}^A}\\
& = \sum_{i>j}\left(R_{ij}^Ap_j-R_{ji}^Ap_i\right)\ln\frac{R_{ij}^Ap_j}{R_{ji}^Ap_i}\\
& \geq 0.
\end{align}
\end{subequations}
The inequality in the last line follows from noting that the difference and log-ratio must have the same sign, so that their product cannot be negative. The same logic holds for the entropy production rate due to transitions coupled to the $B$ reservoir, so that
\begin{equation}
\dot{\Sigma}_B = \dot{S}_B - \frac{\dot{Q}_B}{\kB T_B}\geq 0.
\end{equation}

\section*{Appendix B: Calculational details for information rates in light-harvesting molecular machines}

In Sec.~\ref{PSIIInference} we used the IFAR and EAR to infer information rates in the light-harvesting molecular machines photosystem II and bacteriorhodopsin. We verified these predictions by directly computing information rates in experimentally parameterized stochastic models; here we provide the calculational details.

Light-harvesting machines can be modelled by master equations of the form considered in Appendix A,
\begin{equation}
\frac{\partial}{\partial t} p_i = \sum_j \left[\left(R_{ij}^A + R_{ij}^B\right)p_j - \left(R_{ji}^A + R_{ji}^B\right)p_i\right],
\end{equation}
where the $A$ reservoir is blackbody radiation while the $B$ reservoir is the ambient thermal bath of the cellular environment. For both the 24-state Scheme 4 model for photosystem II detailed in Ref.~\cite[Appendix A.3]{lazar2009approaches} and the 7-state model for bacteriorhodopsin detailed in Ref.~\cite[SI Sec.~IIA]{pinero2024optimization}, each transition is uniquely ascribed to a single reservoir such that for all $i,j$, $R_{ij}^A = 0$ if $R_{ij}^B\neq0$, and vice versa. We parameterize the rate matrices $R_{ij}^A$ and $R_{ij}^B$ for each system according to the experimentally determined rate constants respectively reported in Refs.~\cite{lazar2009approaches} and \cite{pinero2024optimization}.

The information rate due to the dynamics of the light-induced transitions is computed as
\begin{equation}
\begin{aligned}
\dot{I}_A & = \dot{S}_A\\
& = \sum_{i>j}\left[R^A_{ij} \pi_j - R^A_{ji}\pi_i \right] \ln\frac{\pi_j}{\pi_i},
\end{aligned}
\end{equation}
where $\pi_i$ and $\pi_j$ are the steady-state probabilities under the master-equation dynamics.

\bibliography{main}

\end{document}